\documentclass[twocolumn,showpacs,prb,amsmath,amssymb,nofootnoteinbib]{revtex4-2} 
\usepackage[dvipdfmx]{graphicx}
\usepackage{color}
\usepackage{hhline}
\usepackage{mathrsfs}
\usepackage{dcolumn}
\usepackage{bm}% bold math
\usepackage{multirow}
\usepackage{booktabs}
\usepackage{afterpage}
\usepackage{amsmath}
\usepackage{braket}
\usepackage{subfigure}
\usepackage{newtxtext} % font times
\usepackage{txfonts} %数式フォント

\begin{document}

\title{Simplification of the local full vertex in the impurity problem in DMFT and its applications for the nonlocal correlation }

\author{Ryota Mizuno}\email{mizuno@presto.phys.sci.osaka-u.ac.jp}
\author{Masayuki Ochi}
\author{Kazuhiko Kuroki}

\affiliation{ Department of Physics, Osaka University, 1-1 Machikaneyama, Toyonaka, Osaka 560-0043, Japan}

\date{\today}
\begin{abstract}
The two-particle vertex function is crucial for the diagrammatic extensions beyond DMFT for the nonlocal fluctuation. However, estimating the two-particle quantities is still a challenging task. 
In this study, 
we propose a simplification of the local two-particle full vertex 
and,
using the simplified full vertex,
we develop two methods to take into account the nonlocal fluctuation.
We apply these methods to several models and confirm that our methods can capture important behaviors such as the pseudo gap in the DMFT + nonlocal calculation.
In addition, the numerical costs are largely reduced compared to the conventional methods.

\end{abstract}
%\pacs{???}
\maketitle

\section{Introduction}\label{sec:2021-04-26-00-49}
The dynamical mean field theory~(DMFT)~\cite{RevModPhys.68.13}  is one of the most successful approaches for strongly correlated systems.
In DMFT,
the lattice problem is solved non-perturbatively by mapping it onto the Anderson impurity problem,
and 
the local temporal fluctuation is treated correctly.
The great advantage of DMFT is its capability to describe some aspects of Kondo physics or the metal-to-Mott insulator transition,
which are closely related to the local fluctuation.

On the other hand,
the nonlocal spatial fluctuation is not taken into account in DMFT,
which is important for the low dimensional systems or at low temperature.
Hence, 
DMFT can not be used for analysis for  some important  phenomena such as the pseudo gap and the anisotropic superconductivity.
To overcome this weakness, 
several extensions 
of DMFT for describing 
the spatial fluctuation have been suggested.
There are mainly two types of extensions beyond DMFT: the cluster type~\cite{RevModPhys.77.1027} and the diagrammatic type~\cite{RevModPhys.90.025003}. 
The concept of the cluster type extensions is very simple and we do not have to calculate two-particle quantities in the impurity problem.
However, the system size is strictly limited due to the rapid increase of the computational cost with increasing the system size.
Hence, 
the nonlocal correlation taken into account is limited to the short-range one. 
On the other hand,
in diagrammatic extensions, 
we can take a large system size and the long-range nonlocal correlation is taken into account. 
However, we have to estimate the two-particle vertices in the impurity problem. 
Estimating these quantities with an exact impurity solver such as the continuous-time Quantum Monte Carlo method (CT-QMC)~\cite{PhysRevB.72.035122,PhysRevLett.97.076405,PhysRevB.76.235123,PhysRevB.74.155107,doi:10.1143/JPSJ.76.114707}
or
the exact diagonalization (ED)~\cite{PhysRevLett.72.1545,PhysRevB.86.165128}
is still a challenging task, especially in the multiband systems,
although some efforts have been carried out~\cite{PhysRevB.94.125153,10.21468/SciPostPhys.8.1.012}.
In addition, 
we have to perform the calculation regarding the vertices which depend on three frequencies as building blocks in the diagrammatic approaches.
This calculation procedure is numerically expensive and prevents us from analyzing the low temperature regime. 
 
In this study,
we suggest a simplification method for the local full vertex. 
We show that the local full vertex can be approximated in a simple form based on its frequency structure.
By using this approximated form of the local full vertex, 
we develop two methods which are useful for taking into account the nonlocal fluctuation.
One is a method by which we can estimate the two-particle full vertex from the one-particle self-energy. 
This enables us to combine the diagrammatic extensions for the nonlocal fluctuation with any impurity solver
since we can estimate the local full vertex which is a building block of the diagrammatic nonlocal calculation as long as we know the local self-energy.
The other is a simplified version of the dual fermion method~\cite{PhysRevB.77.033101,PhysRevB.79.045133, PhysRevB.90.235132,PhysRevB.97.115150,PhysRevB.98.155117},
where the numerical cost is significantly reduced.

This paper is organized as follows.
In Sec.~\ref{sec:2020-12-25-03-56},
we introduce the models and the two-particle Green's function.
In Sec.~\ref{sec:2021-04-16-23-41},
we show that the local full vertex can be approximated in a simple form.
We describe in Sec.~\ref{sec:2021-04-17-01-33} the novel methods developed in the present study.
Results are shown in Sec.~\ref{sec:2021-04-17-01-36}-Sec.~\ref{sec:2020-12-25-04-06}.
The discussion is presented in Sec.~\ref{sec:2021-04-17-01-37}.
The conclusion is given in Sec.~\ref{sec:2021-04-17-01-38}.

\section{Model and two-particle Green's function}\label{sec:2020-12-25-03-56}
% \subsection{definitions}
We consider the Hubbard model % for multiband systems 
described by the following Hamiltonian.
\begin{align}
  H 
  =&
  \sum_{ij}\sum_{\alpha\beta}t_{ij,\alpha\beta}c^{\dagger}_{i\alpha}c_{j\beta} 
  %-
  %\mu \sum_{i}\sum_{\alpha} n_{i\alpha} \nonumber \\
  %&\hspace{30pt}
  +
  \dfrac{1}{4} \sum_{i} \sum_{\alpha\beta\gamma\lambda} U_{\alpha\beta\gamma\lambda} c^{\dagger}_{i\alpha}c^{\dagger}_{i\lambda}c_{i\gamma}c_{i\beta}, 
   \label{eq:2020-05-12-23-57}
\end{align}
where 
the subscripts with Roman letters indicate unit cells 
and
Greek letters the set of the degrees of freedom of spin, orbital, and site.
$t_{ij,\alpha\beta}$ is the hopping integral 
and 
$U_{\alpha\beta\gamma\lambda}$ is the Coulomb repulsion.
$c_{i\alpha}^{(\dagger)}$ is the annihilation (creation) operator.

\begin{figure*}[t]
  \centering
%  \subfigure[$U/t=6$]
  {\includegraphics[width=140mm,clip]{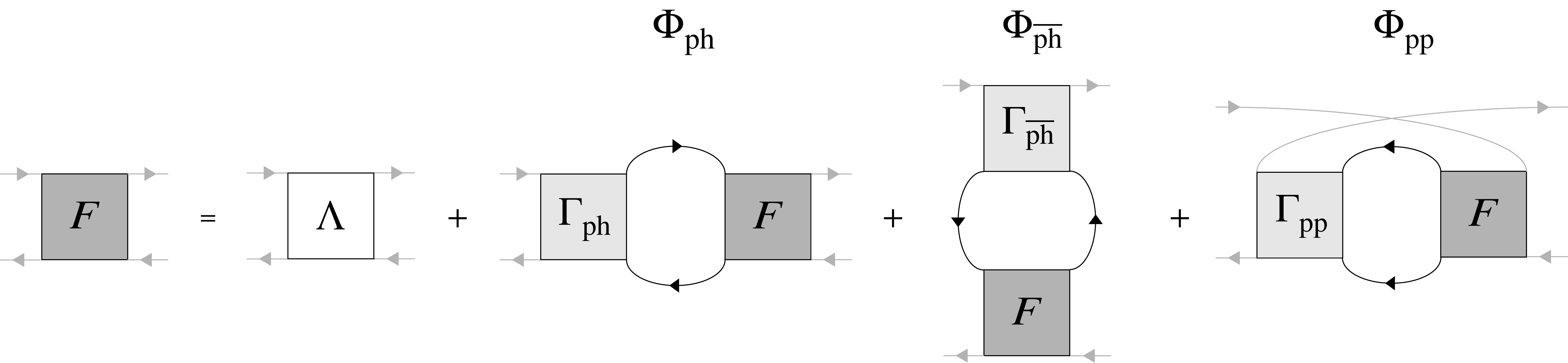}} 
%  \subfigure[$U/t=8$]
%  {\includegraphics[width=140mm,clip]{../graph/Result/graph_2020_04_28/squarex2_delta0.2/Sigma_zenbu_U2Up1J0.5Jp0.5.pdf}} 
  \caption{The decomposition of the full vertex.
    The full vertex can be divided into four parts:
    the fully irreducible part ($\Lambda$) and the reducible parts ($\Phi_{l}$,  $l=$ ph, ${\rm \overline{ph}}$ pp).  
  }
  \label{fig:2020-05-09-22-21}
\end{figure*}

In the presence of the time and lattice translational invariance, 
the two-particle Green's function in the momentum space can be written as 
\begin{align}
  G^{(2)}_{\alpha\beta\gamma\lambda}(\bm{k},\bm{k}',\bm{q}, \tau_{1},\tau_{2},\tau_{3}) 
  =&
  \Bigl< T c_{\bm{k}\alpha}(\tau_{1})c^{\dagger}_{\bm{k}+\bm{q}\beta}(\tau_{2})c_{\bm{k}'+\bm{q}\lambda}(\tau_{3})c^{\dagger}_{\bm{k}'\gamma} \Bigr>,
  \label{eq:2020-10-08-23-02}
\end{align}
where
$c^{(\dagger)}(\tau)=e^{\tau H}c^{(\dagger)}e^{-\tau H}$ is the Heisenberg representation of annihilation (creation) operator.
Fourier transformation is given by 
\begin{align}
  G^{(2)}&(\bm{k},\bm{k}',\bm{q}, \tau_{1},\tau_{2},\tau_{3}) \nonumber \\
  =&
  \dfrac{1}{\beta^{3}} \sum_{nn'm} G^{(2)}(\bm{k},\bm{k}',\bm{q}, i\omega_{n},i\omega_{n'},i\nu_{m})
  e^{-i\omega_{n} \tau_{1}} e^{i(\omega_{n}+\nu_{m})\tau_{2}} e^{-i(\omega_{n'}+\nu_{m})\tau_{3}},
  \label{eq:2020-10-08-23-03}
\end{align}
where 
$\omega_{n}=(2n+1)\pi T$ and  $\nu_{m}=2m \pi T$ with $n, m \in {\mathbb Z}$ are the fermionic and bosonic Matsubara frequencies, respectively.
The two-particle Green's function can be divided into two parts:
disconnected and connected terms.
\begin{align}
  &G^{(2)}_{\alpha\beta\gamma\lambda}(k,k',q) \nonumber \\
  &=
  G_{\alpha\beta}(k) G_{\lambda\gamma}(k') \delta_{q,0} 
  -
  G_{\alpha\gamma}(k) G_{\lambda\beta}(k+q) \delta_{kk'}
  \nonumber \\
  &+
  \hspace{-5pt}\sum_{\alpha'\beta'\gamma'\lambda'}\hspace{-5pt}
  G_{\alpha\gamma'}(k) G_{\lambda'\beta}(k+q) F_{\gamma'\lambda'\alpha'\beta'}(k,k',q) G_{\alpha'\gamma}(k') G_{\lambda\beta'}(k'+q),
\label{eq:2020-10-08-23-04}
\end{align}
where  
$k=(\bm{k},i\omega_{n})$ and $q=(\bm{q},i\nu_{m})$ denote the generalized fermionic and bosonic momentums, respectively.
$F$ is called the full vertex.
To consider the diagrammatic structure of the full vertex $F$, 
we have to define the irreducible susceptibilities concerning the following three channels (${\rm ph, \overline{ph}, pp }$).
\begin{align}
  \chi_{0,\alpha\beta\gamma\lambda}(k,k',q) =& 
  \begin{cases}
    - G_{\alpha\gamma}(k)G_{\lambda\beta}(k+q) \delta_{kk'} \hspace{5pt} &(\text{ph}) \\
    G_{\alpha\beta}(k) G_{\lambda\gamma}(k') \delta_{q0}    &({\overline{\rm ph}}) \\
    G_{\alpha\gamma}(k)G_{\beta\lambda}(-k-q)\delta_{kk'}     &(\text{pp})
  \end{cases}.
  \label{eq:2020-10-08-23-42} 
\end{align}
%ph channel in Eq.~(\ref{eq:2020-10-08-23-42}) is the same as Eq.~(\ref{eq:2020-10-08-23-05}).
The full vertex $F$ can be divided into four parts, 
\begin{align}
  F &= \Lambda + \Phi_{\rm ph} + \Phi_{\rm \overline{ph}} + \Phi_{\rm pp},
  \label{eq:2020-05-09-22-20}
\end{align}
where 
$\Phi_{l}$ $(l={\rm ph,\overline{ph},pp})$
is the set of reducible diagrams in channel $l$,
and 
$\Lambda$ is the set of fully irreducible diagrams. 
The diagrammatic representation is shown in Fig.~\ref{fig:2020-05-09-22-21}.
Since there is no diagram which simultaneously satisfies reducibility  in two or more channels, 
we can write 
\begin{align}
  F =& \Gamma_{l} + \Phi_{l}, 
  \label{eq:2020-05-09-22-53} \\
  \Gamma_{l} =& \Lambda + \Phi_{l_{1}} + \Phi_{l_{2}} \hspace{10pt} (l\neq l_{1} \neq l_{2}), 
  \label{eq:2020-05-09-22-54} \\
  \Phi_{l} =& 
  -\Gamma_{l}\chi_{0} F = -\Gamma_{l}\chi_{l} \Gamma_{l}, 
\end{align}
where 
$\Gamma_{l}$ is the set of diagrams irreducible in channel $l$ 
and 
is called the irreducible vertex in $l$.
The susceptibility in channel $l$ is given by
\begin{align}
  \chi_{l} =& \chi_{0} - \chi_{0}\Gamma_{l}\chi_{l} = \chi_{0} - \chi_{0} F\chi_{0} . \label{eq:2020-05-12-14-40} 
\end{align}
From 
Eqs.~(\ref{eq:2020-05-09-22-53}) to (\ref{eq:2020-05-12-14-40}), which are called the parquet equations~\cite{e_023_03_0489,PhysRevB.86.125114,Janis_1998,PhysRevB.60.11345}, 
we can calculate $F$ exactly if we know the exact $\Lambda$.
However, it is almost impossible to obtain the exact $\Lambda$ 
and the procedure to obtain $\Phi_{l}$ is numerically very expensive. 
Thus, 
some approximations or simplifications have been suggested 
\cite{doi:10.1143/JPSJ.79.094707,PhysRevB.75.165108,PhysRevB.83.035114, PhysRevB.104.035160}.

\section{Simplification of the local full vertex}\label{sec:2021-04-16-23-41}

\begin{figure}[]
  \centering
  {\includegraphics[width=50mm,clip]{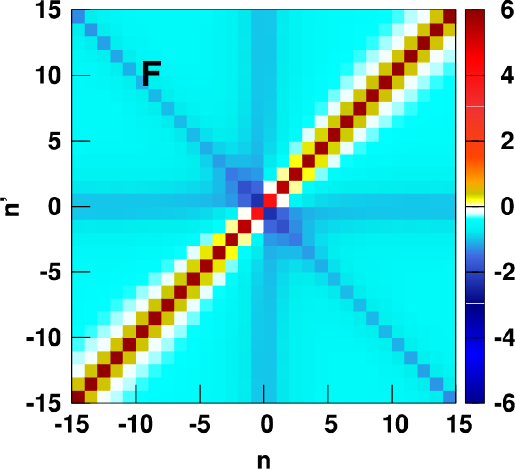}} 
  \caption{The frequency dependence of the full vertex in the charge channel obtained in QMC as an impurity solver.
    The bare interaction $U$ is subtracted.
    The calculations have been performed for the Hubbard model on a square lattice with nearest-neighbor hopping $t$ at $T/t=0.4$, $U/t=5.08$.
    The intensity is given in unit of $4t$.
    This figure is taken from Ref.
    \cite{RevModPhys.90.025003}.
  } 
  \label{fig:2020-06-14-00-39}
\end{figure}

\begin{figure}[]
  \centering
  \subfigure[A diagram independent of $i\omega_{n'}$. ]
  {\includegraphics[width=50mm,clip]{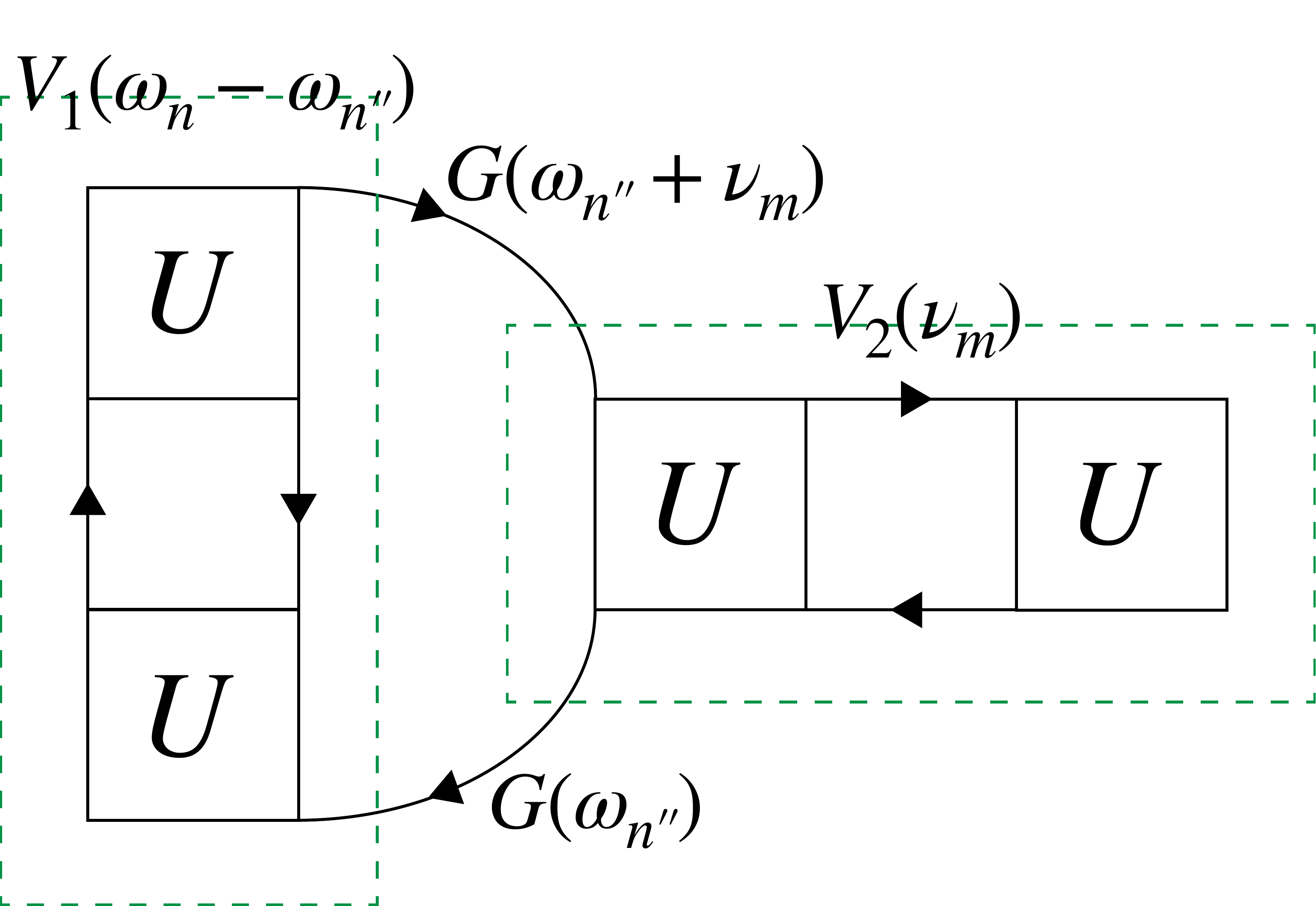}} \hspace{10pt}
  \subfigure[A diagram dependent on $i\omega_{n}$ and $i\omega_{n'}$ independently. ]
  {\includegraphics[width=70mm,clip]{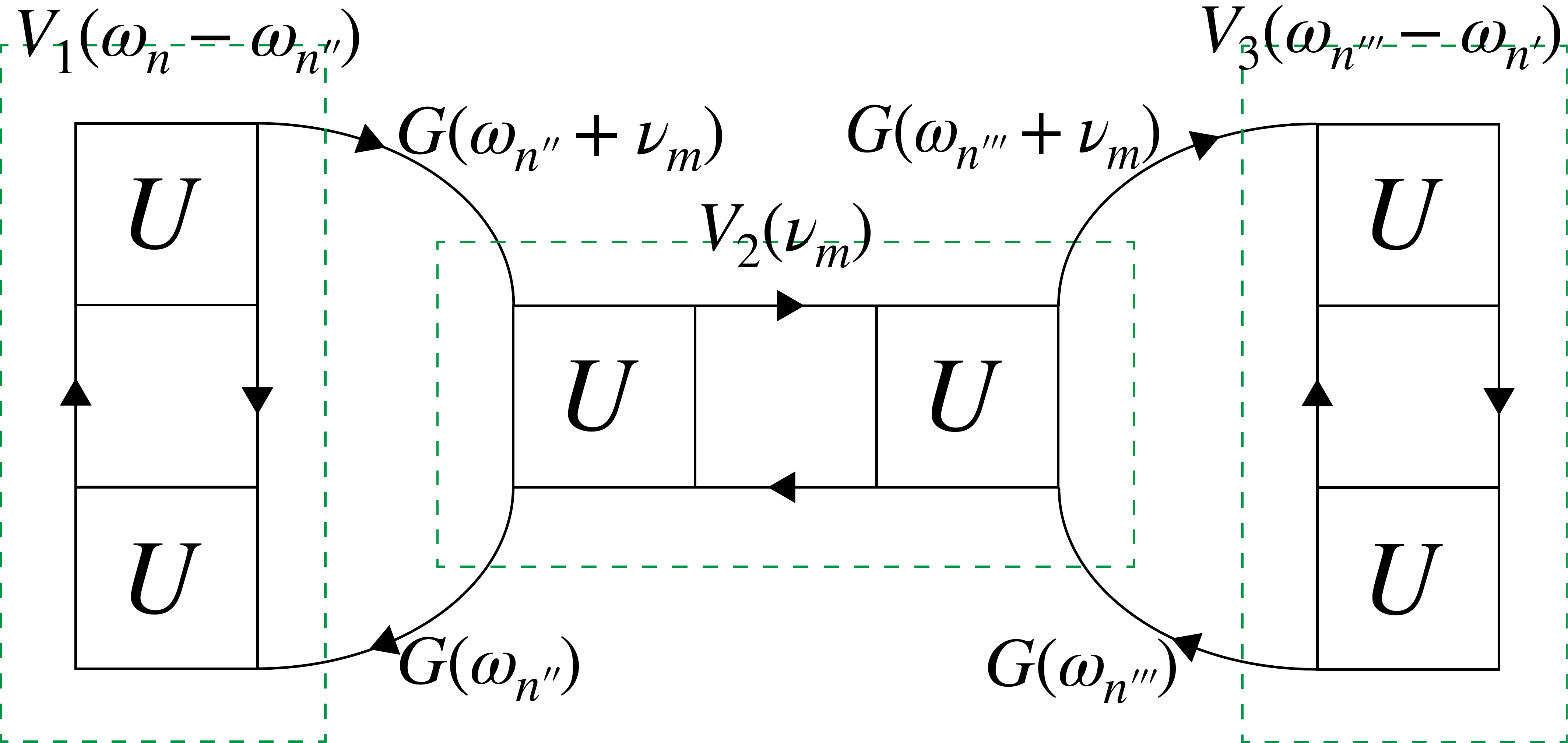}} 
  \caption{Examples of diagrams for each frequency structure.
  The diagrams depicted in (a) and (b) give the cross and  central structures, respectively. } 
  \label{fig:2020-06-14-13-48}
\end{figure}

In this section,
we show that the local full vertex can be approximated in a simple form.
%To this end, 
%we first explain the frequency structure of the full vertex~\cite{RevModPhys.90.025003,PhysRevB.94.235108,PhysRevB.86.125114}.
Figure~\ref{fig:2020-06-14-00-39} shows the full vertex in the charge channel $F^{\rm c}(i\omega_{n},i\omega_{n'},i\nu_{m})$ in the $n$-$n'$ plane calculated with the QMC as the impurity solver~\cite{RevModPhys.90.025003,PhysRevB.94.235108}. 
There are mainly three characteristic structures~\cite{PhysRevB.96.035114,PhysRevB.97.235140,PhysRevB.100.075119,PhysRevB.102.085106,PhysRevB.86.125114,PhysRevB.104.035160}. 
The first one is ``diagonal structure'',
which denotes the large values in the vicinity of the diagonal lines in the $n$-$n'$ plane of 
$F^{\rm c}(i\omega_{n},i\omega_{n'},i\nu_{m})$.
These come from  ${\rm \overline{ph}}$ and pp channels
since 
these channels exhibit large values in the vicinity of $\omega_{n}-\omega_{n'}=0$ and $\omega_{n}+\omega_{n'}+\nu_{m}=0$, respectively.
ph channel takes large values near $\nu_{m}=0$ although it is not depicted in  Fig.~\ref{fig:2020-06-14-00-39}. 
The second one is ``cross structure'',
which denotes the large values in the vicinity of $\omega_{n}=0$ and $\omega_{n'}=0$ lines. 
These come from the combination of some channels depicted in Fig.~\ref{fig:2020-06-14-13-48}~(a), 
which extinguishes the $\omega_{n}$ or $\omega_{n'}$ dependence of $F^{\rm c}(i\omega_{n},i\omega_{n'},i\nu_{m})$ as follows. 
\begin{align}
  T \sum_{n''} V_{1}(\omega_{n}-\omega_{n''})G(\omega_{n''}+\nu_{m})G(\omega_{n''})V_{2}(\nu_{m}).
  \label{eq:2020-06-27-16-10}
\end{align}
Also,
the multiple combinations of some channels depicted in Fig.~\ref{fig:2020-06-14-13-48}~(b)  yields the contribution which depends on $\omega_{n}$ and  $\omega_{n'}$ independently.
\begin{align}
  T^{2}\sum_{n'',n'''}
  &V_{1}(\omega_{n}-\omega_{n''})G(\omega_{n''}+\nu_{m})G(\omega_{n''})
\nonumber \\
  &\times V_{2}(\nu_{m})
  G(\omega_{n'''}+\nu_{m})G(\omega_{n'''})
  V_{3}(\omega_{n'''}-\omega_{n'}).
  \label{eq:2020-06-27-16-11}
\end{align}
These exhibit large values near the center of the $n$-$n'$ plane.
This structure is the third one called  ``central structure''.
As we can see from its origin,
the cross or central structures come from the higher order diagrams than that of the diagonal structure.
Therefore these contributions are important in the strongly correlated regime.

%As mentioned in Sec.~\ref{sec:2021-04-26-00-49},
The complex dependence on the three frequencies of the full vertex makes it difficult to be estimated 
and
brings disadvantages in terms of the numerical costs in the diagrammatic extensions of DMFT, 
in which the calculation is performed using the full vertex as a building block.
To circumvent these difficulties while keeping the important frequency structures~(diagonal, cross, and central), 
we propose a simplification of the full vertex based on the following idea.

$V_{i}(\nu_{m})$ in Eqs.~(\ref{eq:2020-06-27-16-10}) and (\ref{eq:2020-06-27-16-11}) has a large value in the vicinity of $\nu_{m}=0$. 
The values of $V_{i}(\nu_{m})$ in the vicinity of $\nu_{m}=0$ become much larger than others when the two-particle fluctuation expressed by $V_{i}(\nu_{m})$ becomes larger. 
In other words,
$V_{i}(\nu_{m})$ becomes similar to $\delta$ function when the contributions of the three important structures become large.
Considering an extreme case in which  $V_{1}$ and $V_{3}$ are approximated by the $\delta$ function in Eqs.~(\ref{eq:2020-06-27-16-10}) and (\ref{eq:2020-06-27-16-11}),
these vertices can be expressed by the simple product of $V_{2}(\nu_{m})$ and functions which depend on $\omega_{n}, \omega_{n}+\nu_{m},\omega_{n'}$ and $\omega_{n'}+\nu_{m}$.
When we consider channels other than ph in $V_{2}$,
the frequency dependence of $V_{2}$ is replaced as $\nu_{m} \to \omega_{n'}-\omega_{n}$ or $\omega_{n}+\omega_{n'}+\nu_{m}$.
Given this nature of the full vertex, 
we can now approximate the full vertex by employing this simple product form as
\begin{align}
  F(\omega_{n}&,\omega_{n'},\nu_{m}) \nonumber \\
  \approx&
  C_{1}(\omega_{n})C_{2}(\omega_{n}+\nu_{m})F_{0}(\omega_{n},\omega_{n'},\nu_{m})C_{3}(\omega_{n'})C_{4}(\omega_{n'}+\nu_{m}), 
  \label{eq:2021-04-14-21-16} \\
  F_{0}(\omega_{n}&,\omega_{n'},\nu_{m}) \nonumber \\
  =&
  \Lambda + \Phi_{\rm ph}(\nu_{m}) + \Phi_{\rm \overline{ph}}(\omega_{n}-\omega_{n'}) + \Phi_{\rm pp}(\omega_{n}+\omega_{n'}+\nu_{m}).
  \label{eq:2021-04-14-21-17}
\end{align}
The contributions of the constant and diagonal structures are taken into account by $F_{0}$, 
and that of the cross and central structures are treated  by $C_{i}$. 
We also use this idea to interpret that the iterative perturbation theory~(IPT) works well in the vast correlation strength regime despite a perturbation scheme~\cite{PhysRevB.104.035160}.
Here, 
we  regard $\Lambda$ as the bare vertex, namely $\Lambda=U$.
Hence we consider only the bare vertex as the fully irreducible term.
It was shown that the contributions of higher order diagrams of the fully irreducible vertex can become large (even give rise to divergence of irreducible vertices in certain interaction strength) in the previous studies~\cite{PhysRevLett.110.246405,PhysRevB.94.235108}.
However, these contributions were almost invisible in the full vertex.
Therefore, 
we assume that we can ignore the higher order diagrams of the fully irreducible vertex when we estimate only the full vertex.
If we consider the crossing symmetry, we have to take $C_{1}=C_{2}=C_{3}=C_{4}\equiv C$~(see Appendix.~\ref{sec:2021-08-12-15-01} in detail).
The simplification of the full vertex in Eqs.~(\ref{eq:2021-04-14-21-16}) and (\ref{eq:2021-04-14-21-17}) is valid when the two-particle fluctuation is large and $V_{i}$ resembles the $\delta$ function as mentioned above.
On the contrary,
when the two-particle fluctuation is small and $V_{i}$ does not resemble the $\delta$ function, 
this approximation is not good.
However, in such a situation, 
these terms in Eqs.~(\ref{eq:2020-06-27-16-10}) and (\ref{eq:2020-06-27-16-11}) have only a small contribution. 

%Note that we just propose an approximation form of the full vertex 
%and do not propose how to estimate $C$.
%If we can find some way which can estimate $C$,
%it may lead to a new impurity solver.

In the next section,
we propose two methods using this simple product form of the full vertex.
One is the method to estimate $C$ from a given self-energy
and 
the other is a numerically efficient calculation method in the dual fermion system.

\section{Application of simplified full vertex}\label{sec:2021-04-17-01-33}
\subsection{Self-energy to Full vertex (S2F)}\label{sec:2020-07-10-21-40}

Here,
we show that we can estimate the approximate form of the full vertex provided that the self-energy  is already  obtained. 
This enables us to apply diagrammatic extensions of DMFT to impurity solvers in which it is difficult to estimate the two-particle quantities.

First, we assume that the self-energy is already obtained. 
We approximate the full vertex in the form given in Eqs.~(\ref{eq:2021-04-14-21-16}) and (\ref{eq:2021-04-14-21-17}) 
and assume $C_{1}=C_{2}=C_{3}=C_{4}\equiv C$.
The exact expression of the correlation part of self-energy using the full vertex is 
\begin{align}
  \Sigma^{\rm CR}&(\omega_{n}) \nonumber \\ 
  =&
  -T^{2}\sum_{n',m} F(\omega_{n},\omega_{n'},\nu_{m})G(\omega_{n'})G(\omega_{n'}+\nu_{m})  U G(\omega_{n}+\nu_{m}).
  \label{eq:2020-06-29-13-03}
\end{align}
From Eqs.~(\ref{eq:2021-04-14-21-16}) and (\ref{eq:2020-06-29-13-03}), we can obtain 
\begin{align}
  \Sigma^{\rm CR}&(\omega_{n}) 
  =
  C(\omega_{n})X(\omega_{n}), 
  \label{eq:2020-06-29-13-04} \\
  X&(\omega_{n}) %\nonumber \\ 
  =
  -T^{2}\sum_{n',m}C(\omega_{n}+\nu_{m})F_{0}(\omega_{n},\omega_{n'},\nu_{m})C(\omega_{n'}) \nonumber \\
  &\times G(\omega_{n'})C(\omega_{n'}+\nu_{m}) 
  G(\omega_{n'}+\nu_{m})  U G(\omega_{n}+\nu_{m}).
  \label{eq:2020-06-29-13-05} 
\end{align}
Using Eqs.~(\ref{eq:2020-06-29-13-04}) and (\ref{eq:2020-06-29-13-05}), 
we can estimate $C$ in the following steps:
\begin{enumerate}
  \item[(i)] calculate $F_{0}$ by the simplified parquet method~\cite{doi:10.1143/JPSJ.79.094707,PhysRevB.104.035160}. 
  \item[(ii)] calculate $X$ by Eq.~(\ref{eq:2020-06-29-13-05}).  
  \item[(iii)] obtain $C$ by  $C=\Sigma^{\rm CR}X^{-1}$. 
  \item[(iv)] Go back to (ii) (iterate until convergence). 
\end{enumerate}
After convergence, 
we can obtain the full vertex by Eqs.~(\ref{eq:2021-04-14-21-16}) and (\ref{eq:2021-04-14-21-17}).
We will call this method ``S2F~(self-energy to Full vertex)''.
%When we perform EDF calculation using the full vertex obtained by S2F,
%we omit $C_{1}$ and $C_{3}$ in Eq.~(\ref{eq:2020-06-29-15-36}) 
%to avoid an unphysical result
%for the same reason as mentioned in Sec.~\ref{sec:2020-07-10-21-41}.

\begin{figure}[h]
  \centering
%  \subfigure[diagram independent of $i\omega_{n'}$ ]
  {\includegraphics[width=85mm,clip]{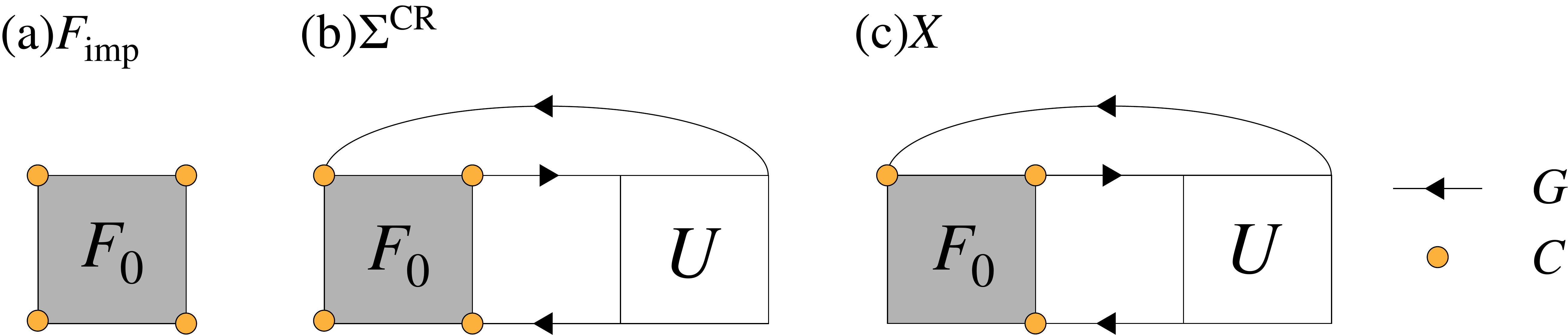}}
  \caption{Diagrammatic representation of (a)~the full vertex, (b)~the correlation part of the self-energy, and (c)~the function $X$ in the S2F method. } 
  \label{fig:2020-06-29-23-35}
\end{figure}

\subsection{Efficient Dual Fermion (EDF)}\label{sec:2020-07-10-21-41}
The dual fermion method~\cite{PhysRevB.77.033101,PhysRevB.79.045133, PhysRevB.90.235132,PhysRevB.97.115150,PhysRevB.98.155117} is one of the extensions of DMFT, in which the nonlocal correlation effect is taken into account by solving the problem in the auxiliary particle system called ``dual fermion system''.
The local full vertex obtained by DMFT is regarded as the bare vertex in the dual fermion system~[see Appendix~\ref{sec:2021-01-01-01-30}].
Although we can adopt the method of the diagram expansion as in the ordinary lattice systems, 
the numerical cost for solving the dual fermion problem can be much higher than that of the ordinary lattice problem even if we employ the same diagrams in each system.
This is because the bare vertex depends on three frequencies in the dual fermion system while it has no frequency dependence in the ordinary lattice system. 

Given the above,
here, we propose a  numerically efficient (low-cost) approximation in the dual fermion method using the simplified form of the full vertex in Eqs.~(\ref{eq:2021-04-14-21-16}) and (\ref{eq:2021-04-14-21-17}).
First, we omit the ${\rm \overline{ph}}$ and pp parts in Eq.~(\ref{eq:2021-04-14-21-17}) for low-cost calculation.
Hence, the full vertex in the local impurity system $F_{\rm imp}$  is approximated as 
\begin{align}
  F_{\rm imp}(\omega_{n}&,\omega_{n'},\nu_{m}) \nonumber \\
  \approx&
  \hspace{5pt}
  C_{1}(\omega_{n})C_{2}(\omega_{n}+\nu_{m})\gamma(\nu_{m})C_{3}(\omega_{n'})C_{4}(\omega_{n'}+\nu_{m}), 
  \label{eq:2020-06-29-14-58}\\
  \gamma(\nu_{m})
  =&
  \Lambda + \Phi_{\rm ph}(\nu_{m}).
  \label{eq:2020-06-29-14-59}
\end{align}
We apply the ladder approximation in the dual fermion system. 
By moving the correction factor $C(\omega_{n})$ from the full vertex $F_{\rm imp}$ to the dual Green's function $G_{\rm dual}$, 
we define the following Green's function-like and susceptibility-like quantities.
\begin{align}
 % \tilde{G}^{C_{i}}(k) =& C_{i}(\omega_{n})G_{\rm dual}(k) 
 % \label{eq:2020-06-29-15-25} \\
  \tilde{G}^{C_{ij}}(k) =& C_{i}(\omega_{n})G_{\rm dual}(k) C_{j}(\omega_{n}),
  \label{eq:2020-06-29-15-26} \\
  \tilde{\chi}_{0}^{C}(q) =& -\sum_{k} \tilde{G}^{C_{13}}(k)\tilde{G}^{C_{42}}(k+q).
  \label{eq:2020-06-29-15-27}
\end{align}
Similarly, the charge and spin susceptibility-like quantities can be obtained as  
\begin{align}
  \tilde{\chi}_{r}^{C}(q) =& \tilde{\chi}_{0}^{C}(q) \bigl[ 1 + \gamma_{r}(\nu_{m}) \tilde{\chi}_{0}^{C}(q) \bigr]^{-1}
  \hspace{10pt} 
  (r = {\rm c,s}) .
  \label{eq:2020-06-29-15-32}
\end{align}
With these, we can obtain the dual self-energy as follows.
\begin{align}
  \tilde{\Sigma}(k)
  =&
  \dfrac{1}{4} \sum_{q} C_{1}(\omega_{n})\Bigl[ V_{\rm c}(q) + 3V_{\rm s}(q) \Bigr]\tilde{G}^{C_{24}}(k+q) C_{3}(\omega_{n}),
  \label{eq:2020-06-29-15-36} \\
  V_{r}(q)
  =&
  2 \gamma_{r}(\nu_{m})\tilde{\chi}_{r}^{C}(q) \gamma_{r}(\nu_{m})  - \gamma_{r}(\nu_{m})\tilde{\chi}_{0}^{C}(q) \gamma_{r}(\nu_{m}). 
  \label{eq:2020-06-29-15-37} 
\end{align}
%As one can see,
In the conventional dual fermion method,
we have to perform the calculation using the local full vertex which depends on three frequencies. 
Hence,
it is difficult to analyze the low temperature regime
since the numerical cost increases as ${\cal O}(N_{\omega}^{3})$ with increasing the required number of Matsubara frequency $N_{\omega}$. % in the low temperature regime.
On the other hand,
in our method here,
%owing to the approximation of Eq.~(\ref{eq:2020-06-29-14-58}),
we do not have to treat all three frequencies practically. 
The numerical cost is reduced as ${\cal O}(N_{\omega}^{3}) \to {\cal O}(N_{\omega})$.
%, where $N_{\omega}$ is the number of the mesh of the frequency.
We will call this method ``efficient dual fermion~(EDF)''.

The local full vertex can be obtained in the form of Eq.~(\ref{eq:2020-06-29-14-58}) by using S2F (see Sec.~\ref{sec:2020-12-25-04-06}).
In the actual S2F + EDF calculation, 
we find that the correction factors $C_{1}$ and $C_{3}$ in Eq.~(\ref{eq:2020-06-29-15-36}), 
which remain without being integrated out, 
leads to unphysical negative values in the spectral function. 
To avoid this problem, 
we omit $C_{1}$ and $C_{3}$ in Eq.~(\ref{eq:2020-06-29-15-36}) in the actual calculation. 
We stress that even if these non-integrated-out corrections in Eq.~(\ref{eq:2020-06-29-15-36}) are omitted, 
the effects of $C_{1}$ and $C_{3}$ 
are taken into account through Eqs.~(\ref{eq:2020-06-29-15-26})-(\ref{eq:2020-06-29-15-32}) and (\ref{eq:2020-06-29-15-37}). 

Another way to obtain the local full vertex in the form of Eq.~(\ref{eq:2020-06-29-14-58}) is to use the IPT + parquet method 
as an impurity solver~\cite{PhysRevB.104.035160}~(see Sec.~\ref{sec:2020-12-24-20-47}). 
In this case, 
we can omit the S2F procedure because the full vertex in IPT + parquet already has the form of Eqs.~(\ref{eq:2021-04-14-21-16}) and (\ref{eq:2021-04-14-21-17}), 
which guarantees the required frequency dependencies,
although $C_{1}=C_{2}=C_{3}=C_{4}$ is not satisfied and hence breaks the crossing symmetry. 
In IPT + parquet, 
the correction factors in Eq.~(\ref{eq:2021-04-14-21-16}) is given as ${C}_{1}(\omega_{n})=[{I}-\hat{B}{\Sigma}^{\rm CR}_{0}(\omega_{n})]^{-1}{A}$
and ${C}_{2}(i\omega_{n})={C}_{3}(i\omega_{n})={C}_{4}(i\omega_{n})={G}_{0}(i\omega_{n}){G}(i\omega_{n})^{-1}$~[see Appendix.~\ref{sec:2021-07-26-21-14} for the details about ${A},\ {B},\ {G}_{0}$, and ${\Sigma}_{0}^{\rm CR}$].
When EDF is combined with IPT + parquet, 
$C_{3}$ in Eq.~(\ref{eq:2020-06-29-15-36}) is omitted in the actual calculation for the same reason as in S2F + EDF, 
but we find that leaving $C_{1}$ does not result in unphysical results, 
and hence $C_{1}$ is not omitted.  
The difference between $C_{1}$ and $C_{3}$ in IPT + parquet is that while the latter 
plays the role of giving the cross and central structures in the frequency dependence of the full vertex, 
the former is a factor  that is related to electron-hole asymmetry~(see Eq.~(26) of Ref.~\cite{PhysRevB.104.035160}).

%****
%In the actual calculation,
%we find that the correction factors which remain without being integrated out such as $C_{1}$ and $C_{3}$ in Eq.~(\ref{eq:2020-06-29-15-36}) give an unphysical result
%when they are obtained in the form $C=\Sigma X^{-1}$ given in Sec.~\ref{sec:2020-07-10-21-40}.
%%[We just know this empirically but can not tell the reason.]
%Therefore,
%when we combine EDF with S2F, 
%we omit $C_{1}$ and $C_{3}$ in Eq.~(\ref{eq:2020-06-29-15-36}) 
%to avoid this problem.
%
%When we use IPT + parquet as an impurity solver~[cite paper1], 
%we can combine EDF with IPT + parquet without S2F procedure,
%since the full vertex in IPT + parquet has the simplified form as in Eqs.~(\ref{eq:2021-04-14-21-16}) and (\ref{eq:2021-04-14-21-17}).
%In this case,
%we omit $C_{3}$ in Eq.~(\ref{eq:2020-06-29-15-36}) because $C_{3}$ has the form $C_{3}=G_{0}G^{-1}$, 
%but
%we leave $C_{1}$ for the following reason.
%$C_{1}$ in Eq.~(\ref{eq:2020-06-29-15-36}) 
%corresponds to the modified factor $(1-B\Sigma_{0})^{-1}A$ in IPT + parquet 
%and 
%this does not give an unphysical result.

\begin{figure}[h]
  \centering
%  \subfigure[diagram independent of $i\omega_{n'}$ ]
  {\includegraphics[width=85mm,clip]{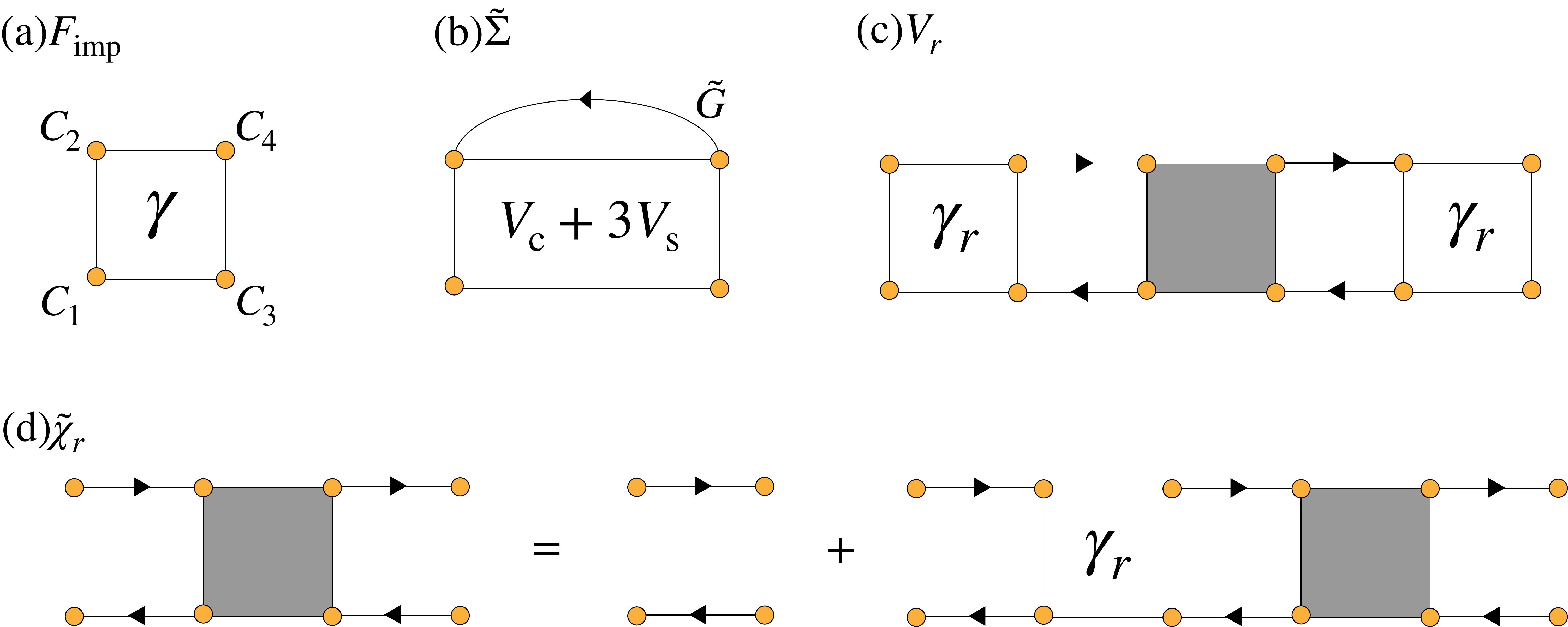}}
  \caption{Diagrammatic representation of 
    (a)~the full vertex in the impurity system, 
    (b)~the correlation part of the self-energy in the dual system, 
    (c)~the function $V_{r}$ , 
    and (d)~$\tilde{\chi}_{r}$
    in the EDF method.
  } 
  \label{fig:2020-06-29-23-35}
\end{figure}

\section{Results of S2F}\label{sec:2021-04-17-01-36}

We study the single-orbital square and cubic lattice models with only the nearest neighbor hopping at half filling 
to compare the S2F results with the numerically exact results in the previous studies~\cite{RevModPhys.90.025003,PhysRevB.86.125114,PhysRevB.96.035114}.
First, we write down the charge and spin channels of $F_{0}$ in Eq.~(\ref{eq:2021-04-14-21-17}) for the following discussion.
\begin{align}
  F_{ 0c}(\omega_{n},\omega_{n'},\nu_{m}) =& 
  U + \Phi_{ c}(\nu_{m}) \nonumber \\ 
  &\hspace{-20pt}- \dfrac{1}{2}[ \Phi_{ c} + 3\Phi_{ s} ](\omega_{n}-\omega_{n'}) \nonumber \\ &\hspace{-20pt} 
  + [ \Phi_{ e} - 3\Phi_{ o} ](\omega_{n}+\omega_{n'}+\nu_{m}),
  \label{eq:2021-07-21-08-57} \\  
  F_{0s}(\omega_{n},\omega_{n'},\nu_{m}) =& 
  -U + \Phi_{ s}(\nu_{m}) \nonumber \\ 
  &\hspace{-20pt}- \dfrac{1}{2}[ \Phi_{ c}  -  \Phi_{ s} ](\omega_{n}-\omega_{n'}) \nonumber \\ &\hspace{-20pt} 
  - [ \Phi_{ e} -  \Phi_{ o} ](\omega_{n}+\omega_{n'}+\nu_{m}), 
  \label{eq:2021-07-21-08-58}   
\end{align} 
where the subscripts $c,s,e$, and $o$ indicate the charge, spin, even, and odd channels, respectively.
The charge and spin (even and odd) channels can be obtained by dividing the ph (pp) channels in terms of the parity of spin~[see Appendix.~\ref{sec:2020-10-09-00-10} in detail].

\begin{figure}[] 
  \centering
  {\includegraphics[clip,width=55mm]{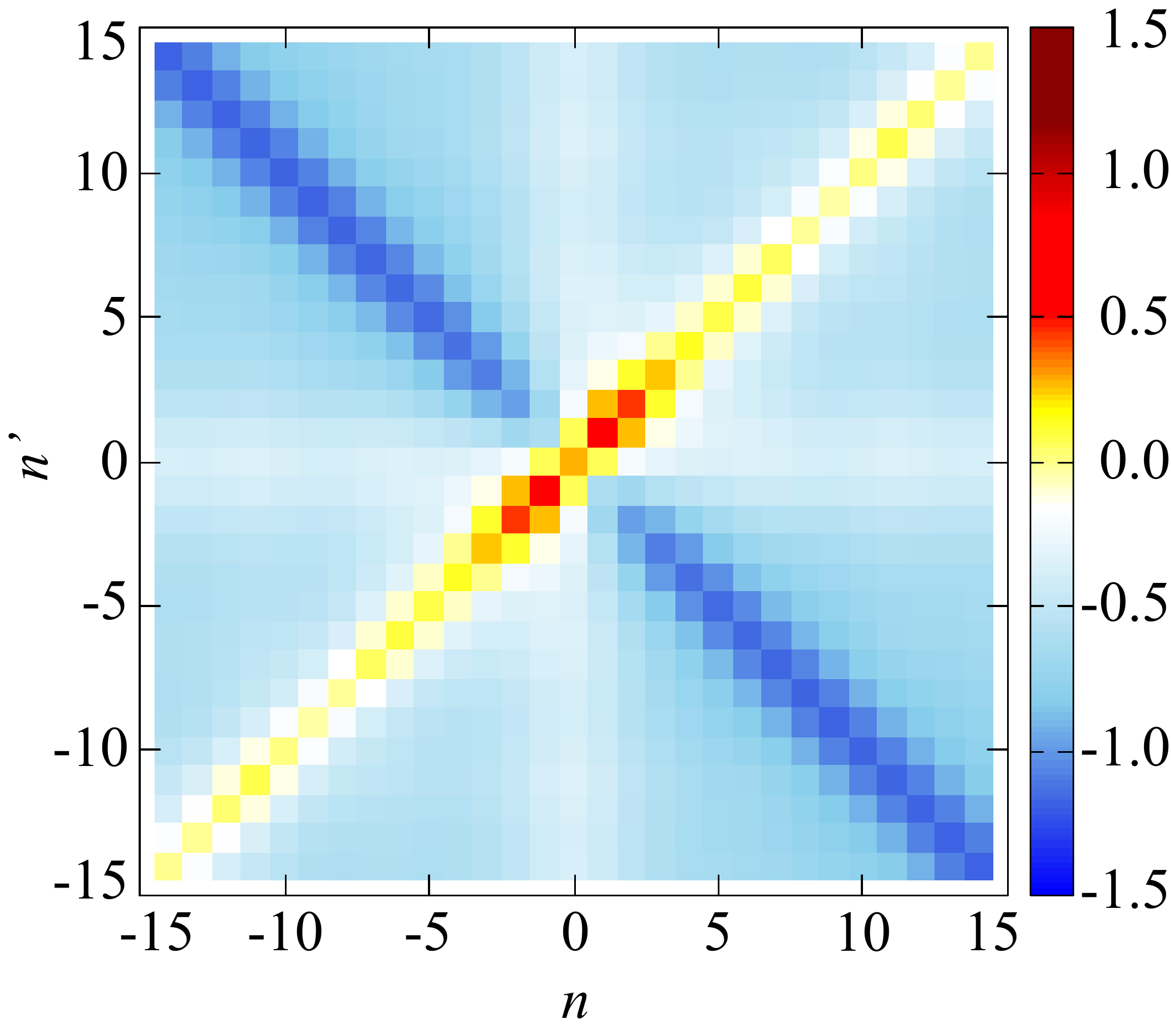}} 
  \caption{
    The full vertex in the charge channel for the square lattice model obtained by S2F. The bare vertex is subtracted as $F_{\rm ch}-U$. 
  } 
  \label{fig:2021-07-19-14-08}
\end{figure}

Figure~\ref{fig:2021-07-19-14-08} shows the full vertex of the square lattice model subtracted by the bare vertex in the charge channel $F_{c}(i\omega_{n},i\omega_{n'},i\nu_{m}) - U$ in the $n$-$n'$ plane obtained by S2F
(The full vertex presented hereafter is always subtracted by the bare vertex). 
The self-energy used in the S2F procedure is obtained by CT-QMC. 
The temperature and the interaction strength are $T/t=0.4$ and $U/t=5.08$, respectively, which are the same as the ones used in the previous study~\cite{RevModPhys.90.025003}.
In Fig.~\ref{fig:2021-07-19-14-08}, 
we can see the three characteristic structures~(diagonal, cross, and central) as in Fig.~\ref{fig:2020-06-14-00-39}.
However, 
the signs of the cross and central structures are different between our result and the previous study~(Fig.~\ref{fig:2020-06-14-00-39})~\cite{RevModPhys.90.025003}.
Namely, 
the values of the cross and central structures are larger than that of the constant background in Fig.~\ref{fig:2021-07-19-14-08} 
while it is smaller in Fig.~\ref{fig:2020-06-14-00-39}.
At present, we are not able to identify the cause for this, 
since the signs of cross and central structures in $n$-$n'$ plane are determined from the complicated relations of vertices.
A possible clue may be the underestimation of the ${\rm \overline{ph}}$ channel $-[\Phi_{\rm c}+3\Phi_{\rm s}]/2$ in Eq.~(\ref{eq:2021-07-21-08-57}),
which we can see from the comparison between the diagonal structures sloping upward in Fig.~\ref{fig:2020-06-14-00-39} and Fig.~\ref{fig:2021-07-19-14-08}.
%It may be one of the causes of the different signs.

\begin{figure}[] 
  \centering
  {\includegraphics[clip,width=85mm]{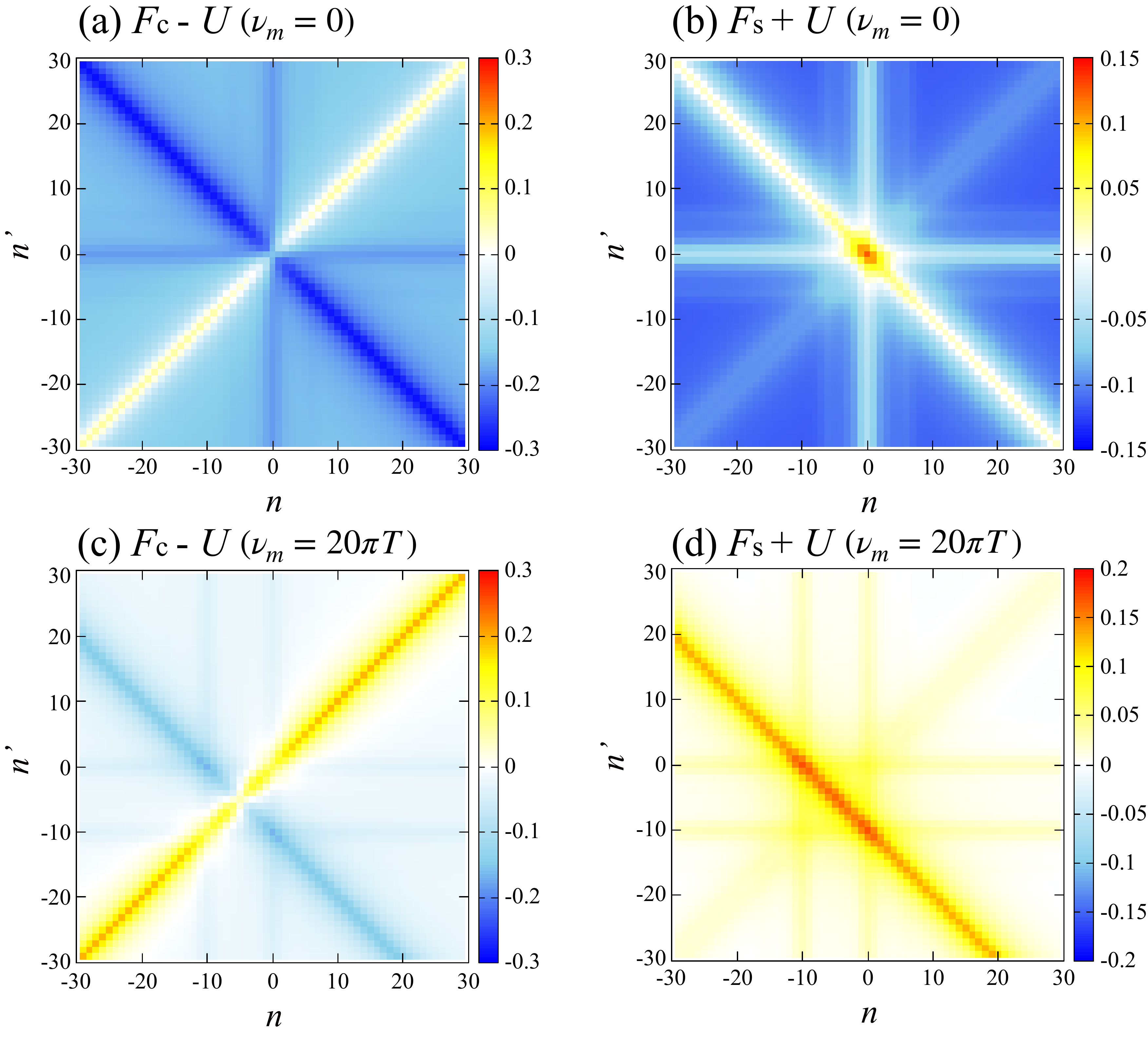}} 
  \caption{
    The full vertex for the cubic lattice model obtained by S2F, where the bare vertex is subtracted as indicated in each figure.
    (a),(c) show the charge channel and (b),(d) show the spin channel.
    (a),(b) show the $n-n'$ plane at $\nu_{m}=0$ and (c),(d) at $\nu_{m}=20\pi T ~(m=10)$. 
    The interaction and temperature are $U/D = 0.5$ and $T/D=1/26$, respectively.
    $D/2=\sqrt{6}t$ is the standard deviation.
  } 
  \label{fig:2021-07-19-14-09}
\end{figure}

\begin{figure}[] 
  \centering
  {\includegraphics[clip,width=85mm]{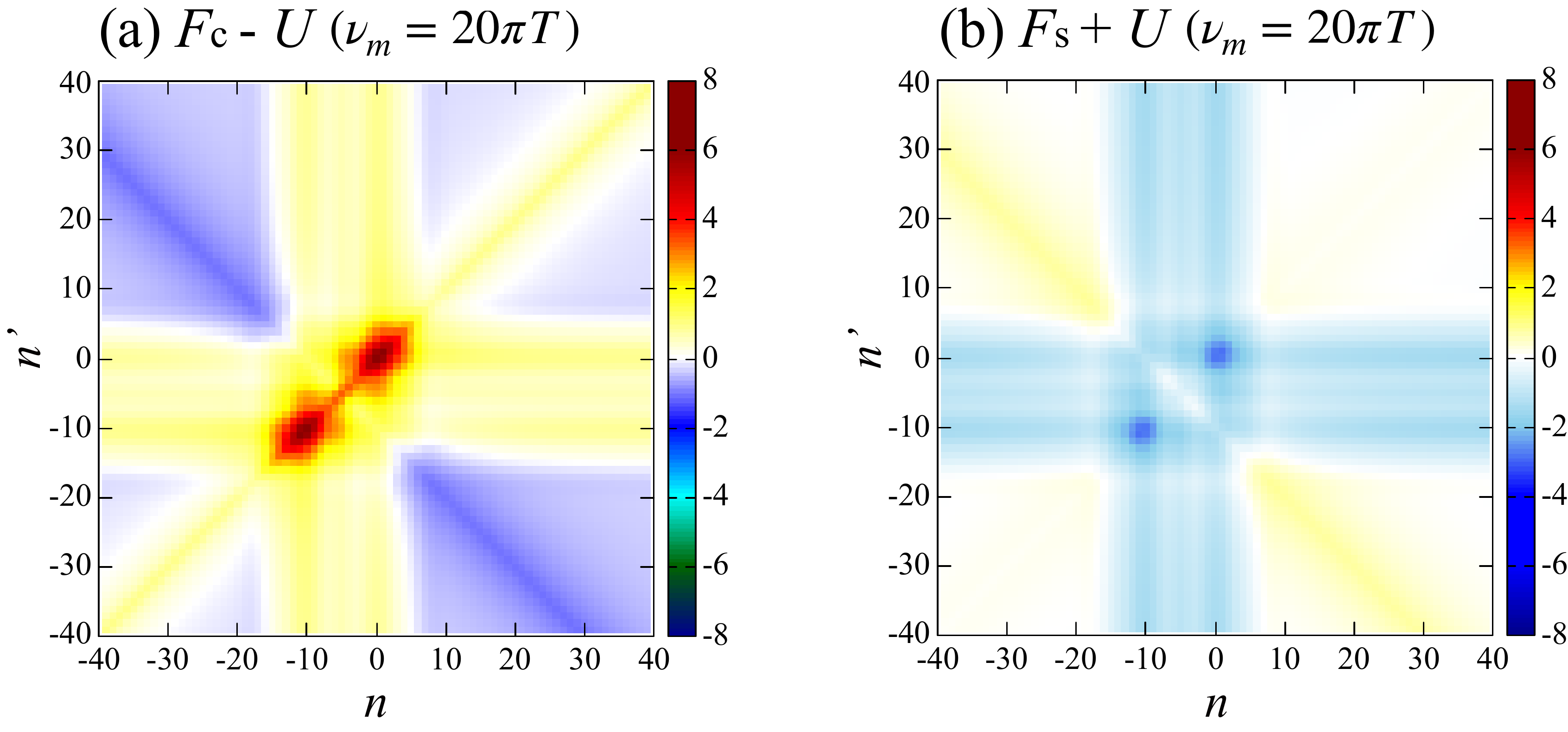}} 
  \caption{
    The full vertex for the cubic lattice model at $\nu_m=20\pi T \ (m=10)$ obtained by S2F, where the bare vertex is subtracted as indicated in each figure.
    (a) and (b) show the charge and spin channels, respectively.
    The interaction and temperature are $U/D = 2$ and $T/D=1/26$, respectively.
    $D/2=\sqrt{6}t$ is the standard deviation.
  } 
  \label{fig:2021-07-23-16-57}
\end{figure}

\begin{figure}[] 
  \centering
  {\includegraphics[clip,width=85mm]{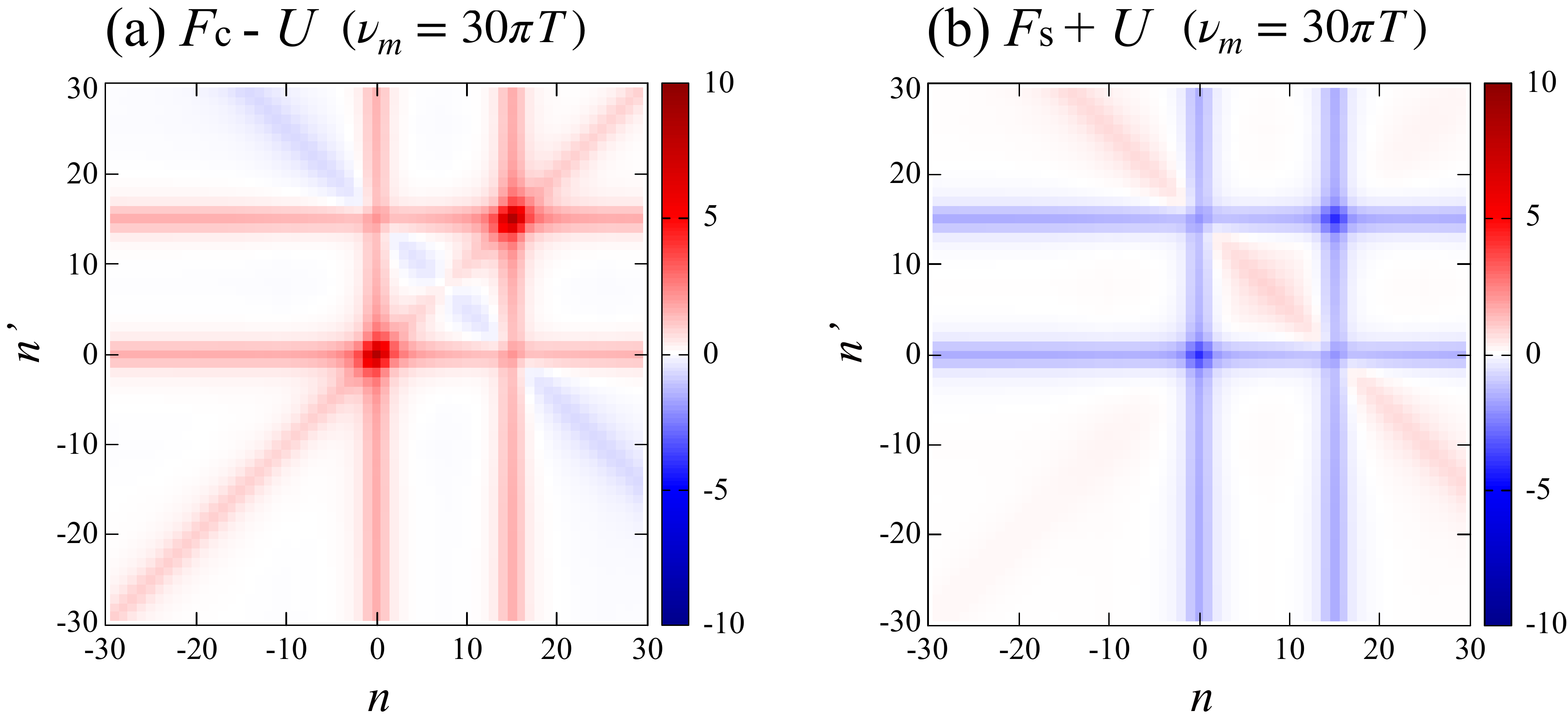}} 
  \caption{
    The full vertex for the cubic lattice model at $\nu_{m}=30\pi T~(m=15)$ obtained by S2F, where the bare vertex is subtracted as indicated in each figure.
    (a) and (b) show the charge and spin channels, respectively.
    The interaction and temperature are $U/D = 2$ and $T/D=1/8$, respectively.
    $D/2=\sqrt{6}t$ is the standard deviation.
  } 
  \label{fig:2021-07-19-14-10}
\end{figure}

Figures~\ref{fig:2021-07-19-14-09} to \ref{fig:2021-07-19-14-10} show the full vertex of the cubic lattice model in the charge and spin channels  obtained by S2F. 
In Fig.~\ref{fig:2021-07-19-14-09}, the interaction strength and temperature are $U/D= 0.5$ and $T/D=1/26$, respectively, 
which are the same as the ones used in Fig.~7 in Ref.~\cite{PhysRevB.86.125114}.
$D/2=\sqrt{6}t$ is the standard deviation of the cubic lattice with only the nearest neighbor hopping.
We can see the three characteristic structures~(diagonal, cross, and central), 
and 
the diagonal structures sloping downward move with increasing the bosonic Matsubara frequency $\nu_{m}$. 
These behaviors are qualitatively consistent with the previous study~\cite{PhysRevB.86.125114}.
However, 
we can see the underestimation of $\Phi_{\rm c}$ and $\Phi_{\rm s}$ also here. 
Especially, in Fig.~\ref{fig:2021-07-19-14-09}~(d), 
where $\nu_{m}=20 \pi T$,
the diagonal structure sloping upward has somewhat weak intensity, 
while it is clearer in the previous study~\cite{PhysRevB.86.125114}.
In Fig.~\ref{fig:2021-07-23-16-57},
we show the results for $\nu_m=20\pi T$, 
with the interaction and temperature taken as 
$U/D=2$ and $T/D=1/26$, respectively,
which are the same as the ones used in Fig.~9 in Ref.~\cite{PhysRevB.86.125114}.
Although there is an underestimation of $\Phi_{\rm c}$ and $\Phi_{\rm s}$,
the frequency structures are qualitatively consistent with the previous study~\cite{PhysRevB.86.125114}.
The large central structure, 
which is clear in Fig.~\ref{fig:2021-07-23-16-57} at present study, 
is unclear in Fig.~9 in Ref.~\cite{PhysRevB.86.125114} because of the large diagonal structure.
However,
we can recognize that the large cross structure indeed exists in both studies.
In Fig.~\ref{fig:2021-07-19-14-10}, 
we take $\nu_m=30\pi T$ with
the interaction and the temperature taken as   $U/D=2$ and $T/D=1/8$, respectively,
which are the same as the ones used in Fig.~5 in Ref.~\cite{PhysRevB.96.035114}.
From these figures, %Fig.~\ref{fig:2021-07-19-14-10} or  Fig.~5 in Ref.~\cite{PhysRevB.96.035114}, 
we can see the structure of the full vertex in a high frequency region since $\nu_{m=10}/D \sim {\cal O}(10)$.
There are sign changes at $\omega_{n}=0$ and $\omega_{n}+\nu_{m}=0$ % , $\omega_{n'}=0$, and $\omega_{n'}+\nu_{m}=0$ 
in Fig.~5 in Ref.~\cite{PhysRevB.96.035114},
while 
there is no sign change 
in Fig.~\ref{fig:2021-07-19-14-10} at present study.
The sign changes can be understood from the discussion in Sec.~\ref{sec:2021-04-16-23-41}.
Considering the extreme case in which $V_{1}$ in  Eqs.~(\ref{eq:2020-06-27-16-10}) or Fig.~\ref{fig:2020-06-14-13-48} is approximated by the $\delta$ function, 
the cross structure is given in the form proportional to $G(\omega_{n})G(\omega_{n}+\nu_{m})$.
Since we consider the electron-hole symmetric case here,
$G(\omega_{n})$ is a pure imaginary function. 
Hence, the cross structure in Fig.~5 in Ref.~\cite{PhysRevB.96.035114} changes its sign at $\omega_{n}=0$ and $\omega_{n}+\nu_{m}=0$,
reflecting the nature of ${\rm Im}G(\omega_{n})$.
On the other hand, 
the cross structure in Fig.~\ref{fig:2021-07-19-14-10} at present study does not change the sign 
because the correction factor $C$, 
which gives the cross structure in the form $C(\omega_{n})C(\omega_{n}+\nu_{m})$, 
is given as $C=\Sigma^{\rm CR}/X$ in Eqs.~(\ref{eq:2020-06-29-13-04}),(\ref{eq:2020-06-29-13-05}) and does not have the Green's function-like frequency dependence.
(see Appendix.~\ref{eq:2021-08-11-21-21}).

The reason why we find qualitative agreement with the previous studies in Fig.~\ref{fig:2021-07-19-14-08} for the square lattice and in Fig~\ref{fig:2021-07-19-14-09}(a)(b) and \ref{fig:2021-07-23-16-57} for the cubic lattice
is because the cross structure is nearly proportional to $G(\omega_{n})G(\omega_{n})$ at small $\nu_{m}$ and the sign change becomes less visible. 
In total,
we can say that S2F works relatively well in the low frequency region, while it becomes less valid in the high frequency regime. 
Despite some inadequacies especially in the high frequency regime, 
we shall see later that this method, combined with EDF, 
can reproduce even quantitatively some physical quantities that were calculated with much larger computational cost.  
An important point that should be stressed here is that, 
despite some insufficiencies  mentioned above, 
there are no other methods, to our knowledge, 
that can take into account, with such low computational cost, 
the cross and central structures in the full vertex, 
which are essential in describing the strong correlation effects. 
In addition, 
as explained later, 
in the dual fermion method, 
the full vertex in the high frequency regime is less important due to the rapid decay of the propagator 
\footnote{
We should note that the discussion here is correct when ${\rm Re}G$ is small and ${\rm Im}G$ is large.
Hence,
S2F results may turn out to be not as good as presented here,
when ${\rm Re}G$ is large and ${\rm Im}G$ is small,
namely,
when the electron-hole asymmetry is large and the correlation strength is small. 
}.

\section{Results of EDF}\label{sec:2020-12-24-20-47}

Here, we show the results of the EDF method.
We employ as the impurity solver the IPT + parquet method, which is an extended version of the iterative perturbation theory~(IPT)~\cite{doi:10.1143/PTPS.46.244, doi:10.1143/PTP.53.970, doi:10.1143/PTP.53.1286, Yamada4, PhysRevB.45.6479,PhysRevLett.77.131,PhysRevB.55.16132,PhysRevB.86.085133,Saso_2001,doi:10.1143/JPSJ.72.777,PhysRevLett.91.156402,Dasari2016} 
developed by the present authors~\cite{PhysRevB.104.035160}, 
and the nonlocal correlation is taken into account by EDF.
In IPT + parquet, 
we can obtain the full vertex in the simplified form given in Eqs.~(\ref{eq:2021-04-14-21-16}) and (\ref{eq:2021-04-14-21-17}) without the S2F procedure.

%the local full vertex already has the simplified form of Eqs.~(\ref{eq:2021-04-14-21-16}) and (\ref{eq:2021-04-14-21-17}), 
%which takes into account the desirable frequency dependencies, 
%although $C_{1}=C_{2}=C_{3}=C_{4}$  is not satisfied and hence breaks the crossing symmetry. 
%Therefore, we can combine IPT + parquet with EDF without the S2F procedure.

We perform two types of calculations: the one-shot and self-consistent calculations.
In the one-shot calculation, we perform only one EDF calculation after the convergence of  local IPT + parquet calculation.
In the self-consistent calculation, IPT + parquet and EDF calculations are repeated alternately until convergence
\footnote{
  The one-particle self-consistency is fulfilled through the self-consistent EDF calculation while the two-particle self-consistency is violated.
  If we extend the simplified parquet method such that it can take into account the effect of the correction factor $C$ as mentioned in Sec.~\ref{sec:2021-08-03-16-22},
  the two-particle self-consistency is also fulfilled.
}
.
Then we use the following update formula for the hybridization function $\Delta(i\omega_{n})$.
\begin{align}
  \Delta^{\rm new}(i\omega_{n})
  =&
  \Delta^{\rm old}(i\omega_{n}) \nonumber \\
  + \hspace{5pt} &
  \xi 
  G^{-1}_{\rm imp}(i\omega_{n})G^{\rm loc}_{\rm dual}(i\omega_{n})[G_{\rm imp}(i\omega_{n})+G_{\rm dual}^{\rm loc}(i\omega_{n})]^{-1}
%  \dfrac{\sum_{\bm{k}}G_{\rm dual}(k)/N_{\bm{k}}}{G_{\rm imp}(i\omega_{n})[G_{\rm imp}(i\omega_{n})+\sum_{\bm{k}}G_{\rm dual}(k)/N_{\bm{k}}]},
  \label{eq:2020-11-19-18-42} 
\end{align}
with 
\begin{align} 
  &G^{\rm loc}_{\rm dual}(i\omega_{n}) = \dfrac{1}{N_{\bm{k}}}\sum_{\bm{k}}G_{\rm dual}(k),
  \label{eq:2020-12-15-16-15}
\end{align}
where $\xi$ is the mixing rate.
Through this formula,
the condition 
\begin{align}
  \sum_{\bm{k}}G_{\rm dual}(k) = 0
  \label{eq:2020-11-20-17-13}
\end{align}
is satisfied~\cite{PhysRevB.79.045133,PhysRevB.90.235132}.
We use the quantity $\alpha_{\rm s}$, which denotes the largest eigen value of $-\gamma_{s}\tilde{\chi}_{0}^{C}$, as a probe of the nonlocal correlation. 
The spin susceptibility ${\chi}_{\rm s}$ diverges when $\alpha_{\rm s}$ reaches 1.

\begin{figure}[] 
  \centering
  {\includegraphics[clip,width=85mm]{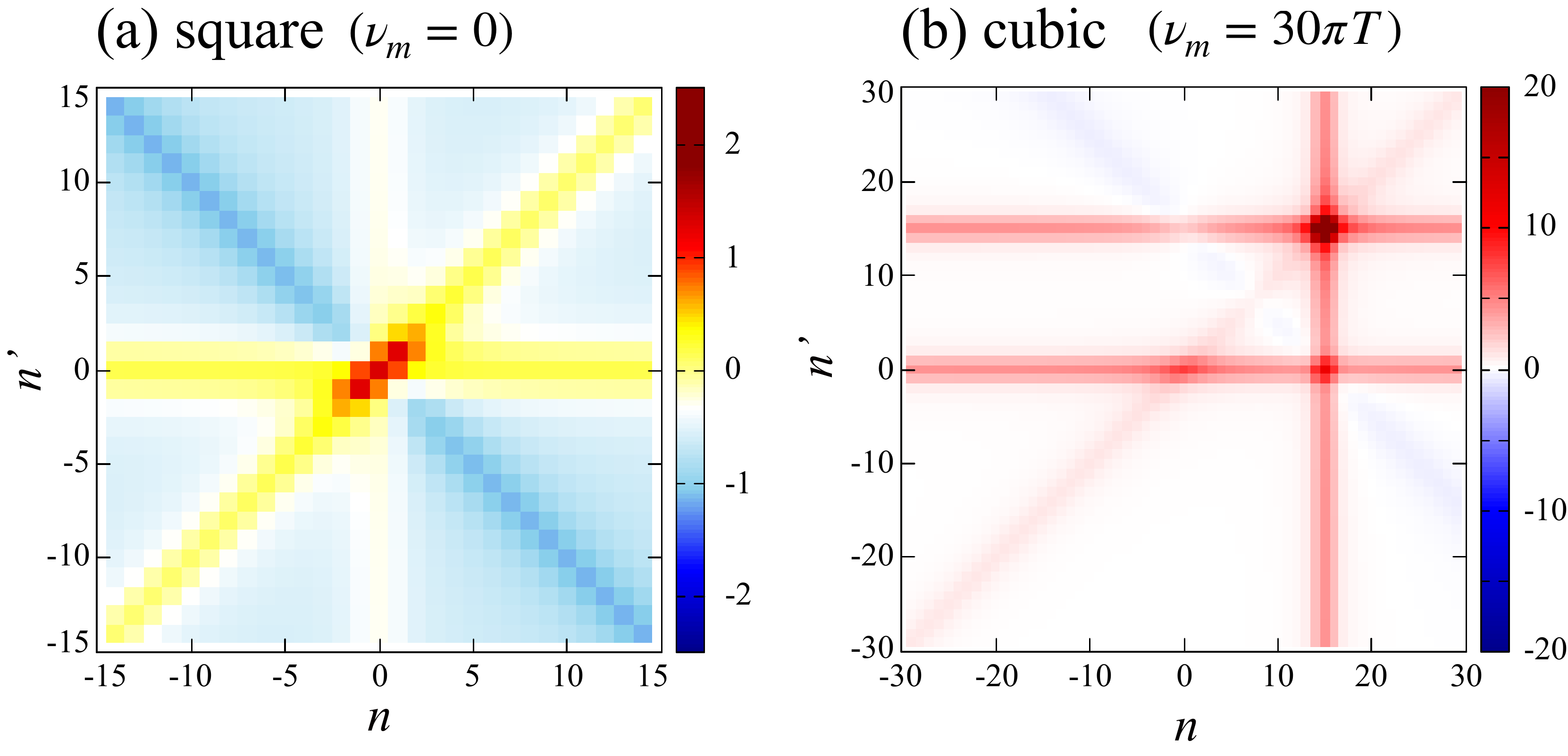}} 
  \caption{
    The full vertex in the charge channel obtained by IPT + parquet, where the bare vertex is subtracted.  
    (a) shows the full vertex of square lattice model at $\nu_{m}=0$. 
    The interaction strength and temperature are $U/t=5.08$ and $T/t=0.4$, respectively.
    (b) shows the full vertex of  cubic lattice model at $\nu_{m}=30\pi T~(m=15)$. 
    The interaction and temperature are $U/D = 2$ and $T/D=1/8$, respectively.
    $D/2=\sqrt{6}t$ is the standard deviation.
  } 
  \label{fig:2021-07-26-17-24}
\end{figure}

Before we move on to the results of EDF, 
we show the full vertex obtained by IPT + parquet. % in Fig.~\ref{fig:2021-07-26-17-24}.
Fig.~\ref{fig:2021-07-26-17-24}~(a) shows the full vertex of the square lattice model in the charge channel at $\nu_{m}=0$.
The interaction strength and temperature are $U/t=5.08$ and $T/t=0.4$, respectively, 
which are the same as those used in Fig.~\ref{fig:2021-07-19-14-08} and  the previous study~\cite{RevModPhys.90.025003}. 
We can see a qualitative agreement with the S2F results~(Fig.~\ref{fig:2021-07-19-14-08}) and the previous study~\cite{RevModPhys.90.025003}. 
Fig.~\ref{fig:2021-07-26-17-24}~(b) shows the full vertex of the cubic lattice model at $\nu_{m}=30 \ (m=15)$ in the charge channel.
The interaction strength and temperature are $U/D=2$ and $T/D=1/8$, respectively, 
which are the same as those used in Fig.~\ref{fig:2021-07-19-14-10} and  the previous study~\cite{PhysRevB.96.035114}.
We can find that the cross structure on the $\omega_{n}=0$ line is absent. 
In IPT + parquet, 
The correction factors in Eq.~(\ref{eq:2021-04-14-21-16}) is given as ${C}_{1}(\omega_{n})=[{I}-\hat{B}{\Sigma}^{\rm CR}_{0}(\omega_{n})]^{-1}{A}$
and ${C}_{2}(i\omega_{n})={C}_{3}(i\omega_{n})={C}_{4}(i\omega_{n})={G}_{0}(i\omega_{n}){G}(i\omega_{n})^{-1}$~(see Appendix.~\ref{sec:2021-07-26-21-14} for the details).
As mentioned in Sec.~\ref{sec:2020-07-10-21-41}, 
${C}_{1}(i\omega_{n})$ is related to the electron-hole asymmetry 
and 
the other three give the cross and central structures.
In the electron-hole symmetric case,
${C}_{1}(i\omega_{n})=1$ and hence the cross structure on the $\omega_{n}=0$ line is absent.
However, 
we consider that the absence of the cross structure on the $\omega_{n}=0$ line at high frequency region hardly matters 
since, 
in the dual fermion calculation, 
the information of the full vertex in the low frequency region  is much more important than that in the high frequency region due to the rapid decay of the propagator in the dual fermion system.

\subsection{Single-orbital model}\label{sec:2021-03-19-14-19}

We begin with the single-orbital square lattice model. %, which is the same model as in Sec.~\ref{sec:2020-11-08-15-30}.  
Here,
we take $32\times 32$ $k$-meshes and 4096 Matsubara frequencies.

\begin{figure}[] 
  \centering
  {\includegraphics[clip,width=75mm]{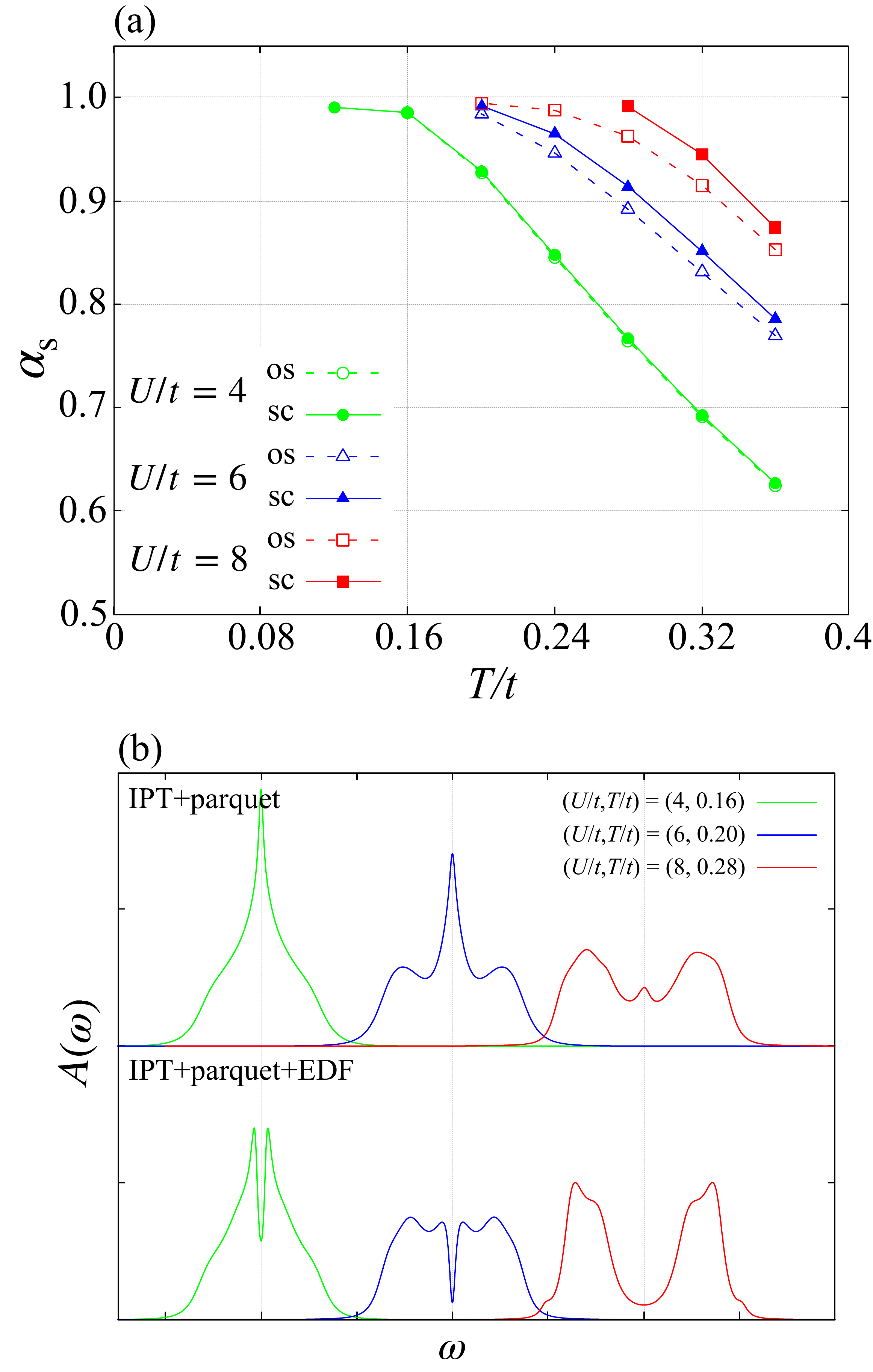}} 
  \caption{
    (a)~The temperature dependence of $\alpha_{\rm s}$ of the single-orbital square lattice model obtained from the IPT + parquet + EDF calculation.
    Green, blue, and red lines indicate the results at  $U/t=4,6$, and $8$, respectively.
    Open symbols indicate the results of the one-shot~(os) calculations and closed symbols the self-consistent~(sc) calculations.  
    (b)~The spectral function $A(\omega)$ of the single-orbital square lattice model obtained from IPT + parquet and IPT + parquet + EDF~(self-consistent)
    Green, blue, and red lines indicate the results at $(U/t, T/t)=(4,0.16), (6,0.20)$, and $(8,0.28)$, respectively.
  } 
  \label{fig:2021-03-19-12-09}
\end{figure}

First,
we study an electron-hole symmetric case,
namely, 
we fix the band filling at $n=0.5$~(half filling) and consider only the nearest neighbor hopping.
Figure~\ref{fig:2021-03-19-12-09}~(a) shows the temperature dependence of $\alpha_{\rm s}$.
In both one-shot and self-consistent calculations, 
$\alpha_{\rm s}$ increases with lowering the temperature or with increasing the interaction strength.
This tendency is  consistent with the previous CT-QMC + DF study~\cite{PhysRevB.90.235132},
where DF is the ordinary  dual fermion calculation.
However, the spin fluctuation in IPT + parquet + EDF seems to be larger than that in CT-QMC + DF. 
A possible reason is as follows, 
although we can not make a simple comparison of $\alpha_{\rm s}$ here and $\lambda_{\rm sp}$ in Ref.~\cite{PhysRevB.90.235132} since they are not exactly the same quantity.
%As mentioned in the previous sections, 
As shown in our separated work~\cite{PhysRevB.104.035160},
in the~(local) impurity problems, 
the results of IPT + parquet and CT-QMC agrees so well that the relative magnitude of physical quantities can be reversed depending on parameters such as the band filling $n$, the temperature $T$, the interaction strength $U$, and so on.
In the EDF procedure, 
at first, 
we ignore the ${\rm \overline{ph}}$ and pp channels in the local impurity full vertex $F_{\rm imp}$.
Namely, 
we regard $\gamma = F_{\rm imp} - ({\rm \overline{ph}} \text{ and pp terms})$ as the bare vertex in the dual fermion system.
Since the ${\rm \overline{ph}}$ and pp channels suppress the spin fluctuation in $F_{\rm imp}$,
we overestimate the spin fluctuation at this step.
Next, 
we apply the ladder approximation in the dual fermion system.
This is the same as the FLEX approximation in ordinary lattice problems, 
and as is well known,
this approximation overestimates the spin fluctuation.
Therefore,
the spin fluctuation is doubly overestimated in IPT + parquet + EDF while it is overestimated only in the ladder approximation in the previous CT-QMC + DF study~\cite{PhysRevB.90.235132}.
Actually, 
there may be one more overestimation in IPT + parquet + EDF.
Since the IPT + parquet results agree very well with the CT-QMC results despite underestimating the diagonal structure~($\Phi_{ c}$, $\Phi_{ s}$), 
it may be that the cross and central structures given by the correction factor $C$ compensate the differences.
The effect of the underestimation of $\Phi_{ c}$ or $\Phi_{ s}$ mentioned in Sec.~\ref{sec:2021-04-17-01-36} may be hidden behind these overestimations. 
Figure~\ref{fig:2021-03-19-12-09}~(b) shows the spectral functions obtained from IPT + parquet and IPT + parquet + EDF~[self-consistent calculation] at the temperature where the spin fluctuation is large~($\alpha_{\rm s} \sim 1$).
Similarly to  previous studies~\cite{doi:10.1143/JPSJ.75.054713,PhysRevLett.102.206401},
we can see the pseudo gap structure caused by the spatial fluctuation in each interaction strength in IPT + parquet + EDF calculation
while it is not seen in the IPT + parquet~(local) calculation.

%\onecolumngrid

\begin{figure}[] 
  \centering
  {\includegraphics[width=90mm,clip]{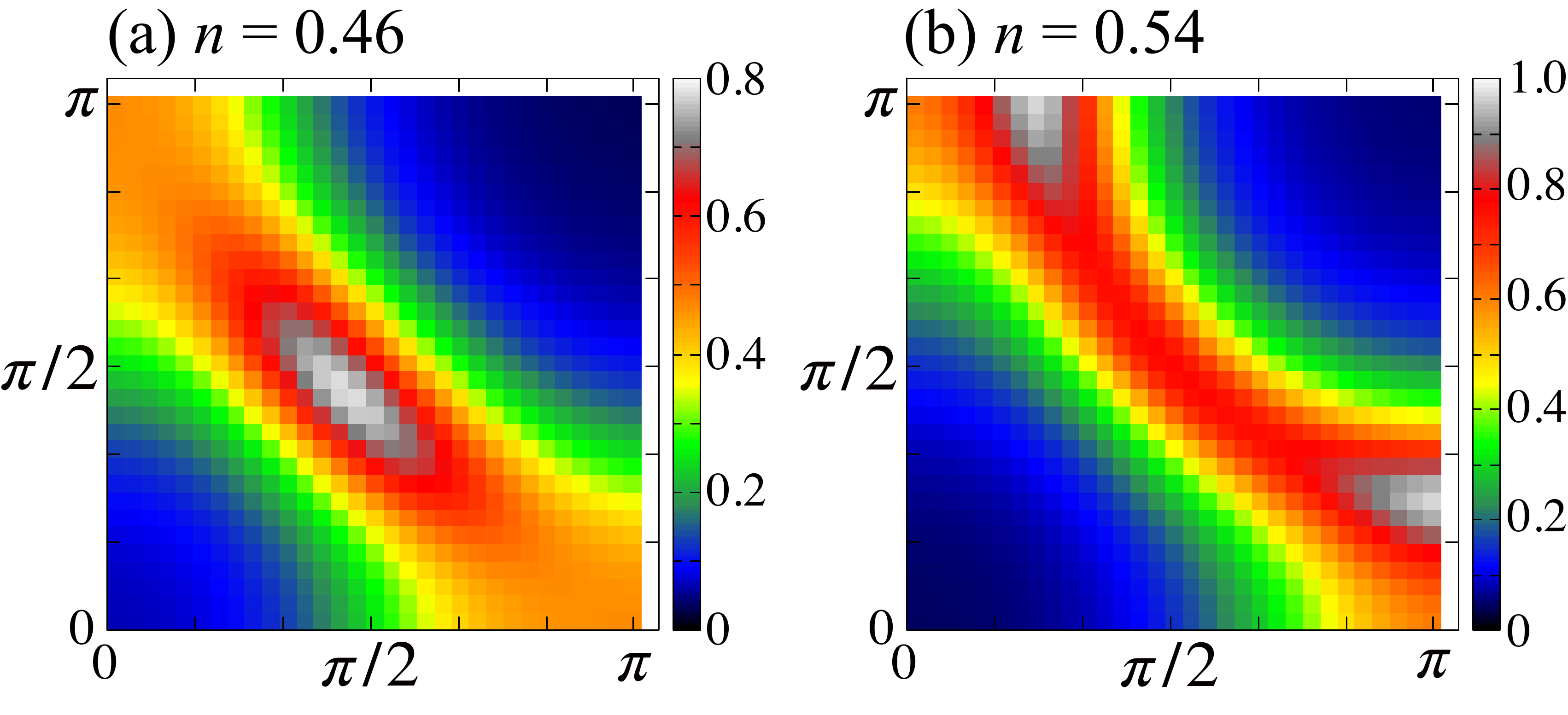}} 
  \caption{
    The quantity $-{\rm Im}G(\bm{k},i\omega_{n=0})/\pi$ of the single-orbital square lattice model obtained by IPT + parquet + EDF~(one-shot calculation) 
    at the band filling (a)~$n=0.46$ and (b)~$n=0.54$.
    The interaction strength and the temperature are  $U/t=8$ and $T/t=0.12$, respectively.
    The hoppings are $t'/t=-0.20, t''/t=0.16$.
  } 
  \label{fig:2021-01-17-14-31}
\end{figure}

Next, we study an electron-hole asymmetric case.
We set $t'/t=-0.20,~t''/t=0.16$, where $t',t''$ are the second and third nearest neighbor hoppings, assuming the single-layer cuprates.
The interaction strength and the temperature are fixed at $U/t=8$ and $T/t=0.12$, respectively.
Figure~\ref{fig:2021-01-17-14-31} shows the quantity $-{\rm Im}G(\bm{k},i\omega_{n=0})/\pi$, which roughly corresponds to the spectral function at the Fermi level, obtained by IPT + parquet + EDF~(one-shot calculation). 
We can see that $-{\rm Im}G(\bm{k},i\omega_{n=0})/\pi$ is suppressed at $(\pi,0), (0,\pi)$ in the hole-dope side~($n=0.46$) and at $(\pi/2,\pi/2)$ in the electron-dope side~($n=0.54$).
This is consistent with the experimental results~\cite{d845815e89a94214b3c7398b7a34611f,PhysRevB.74.224510}
and 
the previous cluster DMFT study, in which the exact diagonalization is used as an impurity solver~\cite{PhysRevB.73.165114}.

\subsection{Two-orbital model}

Next, we study the two-orbital square lattice model with only the intraorbital nearest neighbor hopping.
The one body part of the Hamiltonian is expressed as 
\begin{align}
  H_{0} 
  =&
  \sum_{ij}\sum_{\alpha\beta}t_{ij,\alpha\beta}c^{\dagger}_{i\alpha}c_{j\beta} - \mu \sum_{i}\sum_{\alpha}n_{i\alpha}.
  \label{eq:2020-07-12-20-15}
\end{align}
The interaction part of the Hamiltonian is expressed as 
\begin{align}
  H_{\rm int}
  =&
  \sum_{l} U n_{l \uparrow}n_{l\downarrow} 
  +
  \sum_{l_{1}\neq l_{2}} \sum_{\sigma_{1}\sigma_{2}} U' n_{l_{1}\sigma_{1}} n_{l_{2}\sigma_{2}}
  \nonumber \\
  &+
  \sum_{l_{1}l_{2}} J \bm{S}_{l_{1}} \cdot \bm{S}_{l_{2}} 
  +
  \sum_{l_{1}l_{2}} J' c^{\dagger}_{l_{1}\uparrow}c^{\dagger}_{l_{2}\downarrow} c_{l_{2}\downarrow}c_{l_{2}\uparrow},
  \label{eq:2020-07-02-21-25}
\end{align}   
where 
the degrees of freedom of orbital are expressed by $l$ and spin by $\sigma$.
$U^{(\prime)}$ is the intraorbital~(interorbital) interaction, 
and 
$J$ and $J'$ represent the Hund's coupling and pair hopping, respectively. 
Then, 
the interaction matrices in the charge and spin channels are expressed as
\begin{align}
  \Bigl( U^{\rm c}_{l_{1}l_{2}l_{3}l_{4}}, U^{\rm s}_{l_{1}l_{2}l_{3}l_{4}} \Bigr)
  =
  \begin{cases}
    (U, U) \hspace{20pt} &(l_{1}=l_{2}=l_{3}=l_{4}) \\
    (2U'-J, J) \hspace{20pt} &(l_{1}=l_{2}\neq l_{3}=l_{4}) \\
    (2J-U', U') \hspace{20pt} &(l_{1}=l_{3}\neq l_{2}=l_{4}) \\
    (J', J') \hspace{20pt} &(l_{1}=l_{4}\neq l_{2}=l_{3})
  \end{cases}.
  \label{eq:2020-07-02-21-39}
\end{align}
We set $t_{1}=t_{2}=t$, where $t_{\alpha}=t_{i,i+1, \alpha\alpha}$ is the nearest neighbor hopping of orbital $\alpha$ and $t$ is the unit of energy.
The onsite energy difference is $\delta=t_{ii,11}-t_{ii,22}$, 
and 
the interactions are $U'=U-2J, J=J'=U/4$.
We take $32\times 32$ $k$-meshes and 4096 Matsubara frequencies.
Here we fix the onsite energy difference $\delta/t=1.6$, and the band filling at $n=1.1$.
We intentionally avoid the calculation results at half-filling, which turns out to require special care due to spontaneous symmetry breaking.
We plan to present these results in future publications.

\begin{figure}[] 
  \centering
  {\includegraphics[clip,width=75mm]{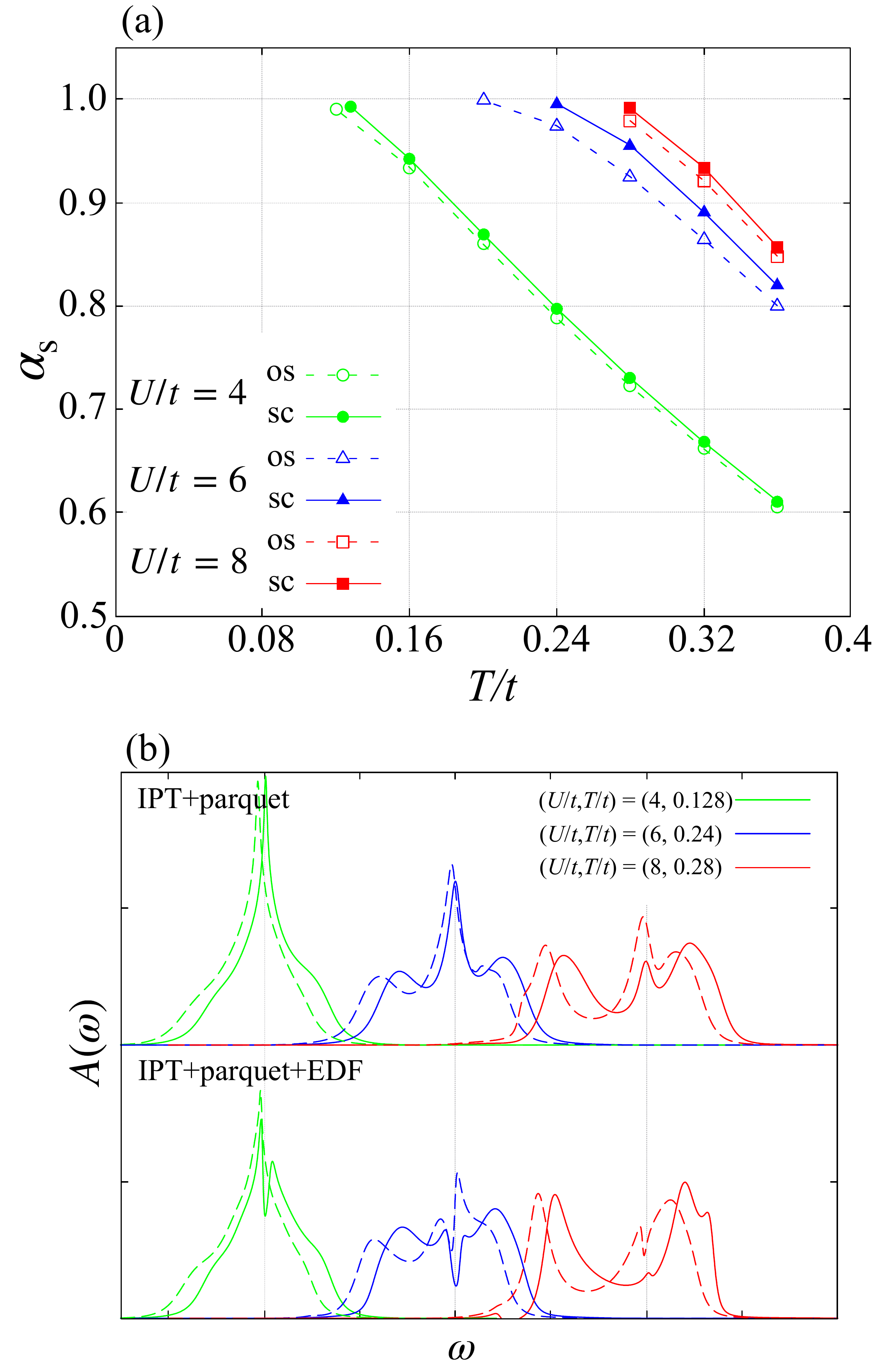}} 
  \caption{(a)~The temperature dependence of $\alpha_{\rm s}$ of the two-orbital square lattice model obtained from the IPT + parquet + EDF calculation.
    Green, blue, and red lines indicate the results at  $U/t=4,6$, and $8$, respectively.
    Open symbols indicate the results of the one-shot~(os) calculations and closed symbols the self-consistent~(sc) calculations.  
    (b)~The spectral function $A(\omega)$ of the two-orbital square lattice model obtained from IPT + parquet and IPT + parquet + EDF~(self-consistent)
    Green, blue, and red lines indicate the results at $(U/t, T/t)=(4,0.128), (6,0.24)$, and $(8,0.28)$, respectively.
    Solid and dashed lines indicate the orbitals 1 and 2, respectively.
  } 
  \label{fig:2021-03-19-12-10}
\end{figure}

Figure~\ref{fig:2021-03-19-12-10}~(a) shows the temperature dependence of $\alpha_{\rm s}$ of the two-orbital square lattice model for several interaction strengths.
Similarly to the single-orbital case,
$\alpha_{\rm s}$ increases with lowering the temperature or with increasing the interaction strength 
in both the one-shot and self-consistent calculations.
However, unlike the single-orbital case,
the difference between the one-shot and self-consistent calculations is largest at $U/t=6$ and is not monotonic in terms of $U$.
In the situation considered here, 
the band fillings of two orbitals are not fixed at half-filling, where the spin fluctuation becomes largest, 
and are determined by the on-site energy difference $\delta$ and the Hartree-Fock term in the self-energy.
Hence,
the $U$ dependence of the spin fluctuation is more complex than that of the single-orbital case in Sec.~\ref{sec:2021-03-19-14-19}.

Figure~\ref{fig:2021-03-19-12-10}~(b) shows the spectral functions obtained from IPT + parquet and IPT + parquet + EDF~[self-consistent calculation]
at the temperature where the spin fluctuation is large~($\alpha_{\rm s} \sim 1$).
Also here,
we can see the pseudo gap behavior caused by the spin fluctuation at each interaction strength.

%\clearpage
\subsection{Bilayer model}

Here, we study the bilayer model on the square lattice.
The Hamiltonian of this model is expressed as 
\begin{align}
  H
  =&
  \sum_{\braket{ij}}\sum_{\alpha}t c^{\dagger}_{i\alpha}c_{j\alpha} + \sum_{i}\sum_{\alpha\neq\beta}t_{\perp}c^{\dagger}_{i\alpha}c_{i\beta} 
  +
  \sum_{i}\sum_{\alpha}Un_{i\alpha}n_{i\alpha},
  \label{eq:2020-07-12-21-39}
\end{align}
%Fig. \ref{fig:2020-07-02-15-38} shows a schematic of the bilayer model.
where $t~(t_{\perp})$ represents the intralayer~(interlayer) hopping and $U$ the on-site interaction,
The temperature is fixed as $T/t=0.2$, 
and the hopping ratio $t_{\perp}/t=1.0$.
We take $32\times 32$ $k$-meshes and 4096 Matsubara frequencies.
Since the two sites are equivalent in this model,
we show only the quantities of site 1 and omit the site index.

\begin{figure}[] 
  \centering
  {\includegraphics[clip,width=75mm]{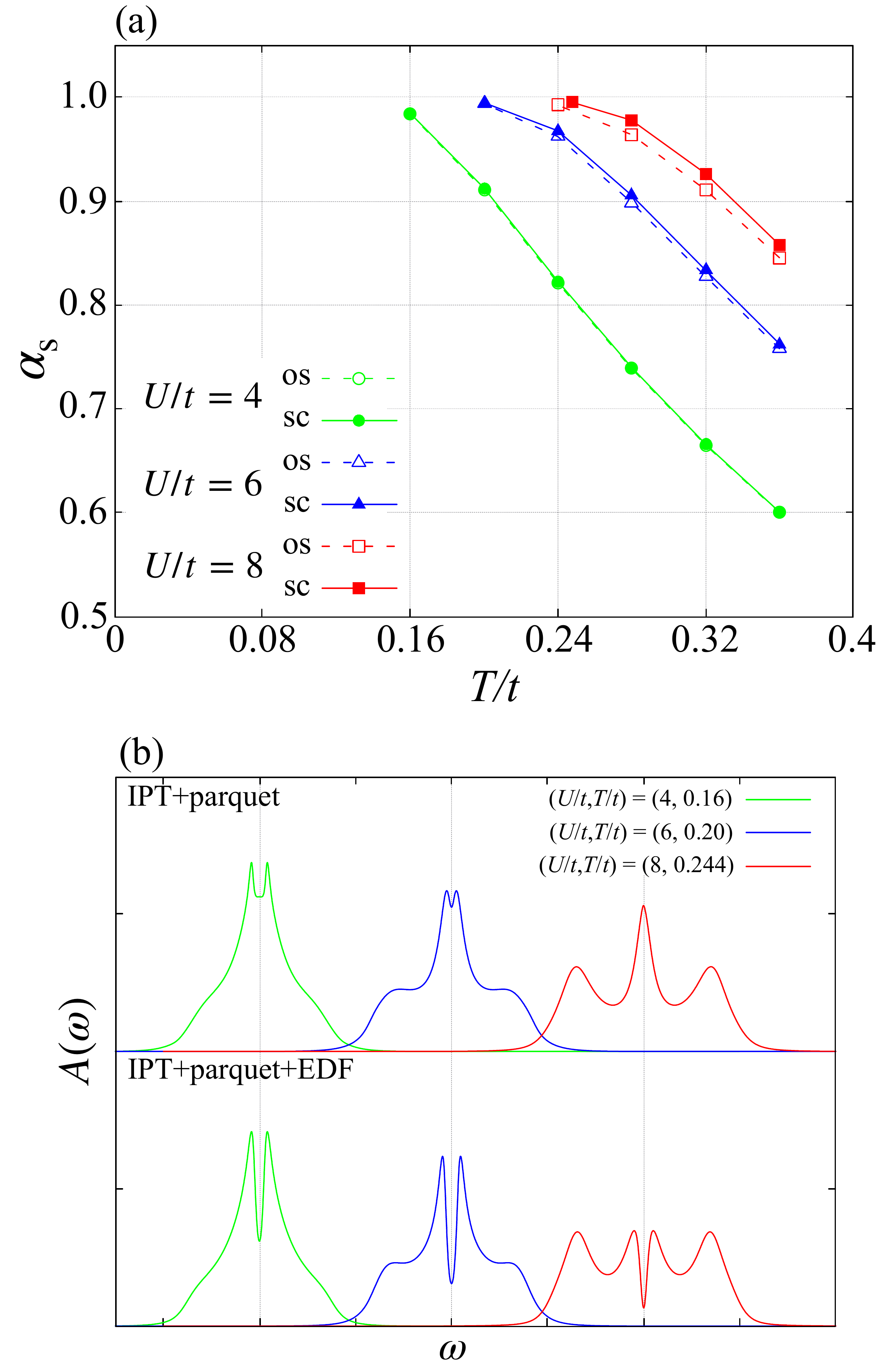}} 
  \caption{(a)~The temperature dependence of $\alpha_{\rm s}$ of the bilayer model obtained from the IPT + parquet + EDF calculation.
    Green, blue, and red lines indicate the results at  $U/t=4,6$, and $8$, respectively.
    Open symbols indicate the results of the one-shot~(os) calculations and closed symbols the self-consistent~(sc) calculations.  
    (b)~The spectral function $A(\omega)$ of the bilayer model obtained from IPT + parquet and IPT + parquet + EDF~(self-consistent)
    Green, blue, and red lines indicate the results at $(U/t, T/t)=(4,0.16), (6,0.20)$, and $(8,0.244)$, respectively.
  } 
  \label{fig:2021-03-19-12-11}
\end{figure}

Figure~\ref{fig:2021-03-19-12-11}~(a) shows the temperature dependence of  $\alpha_{\rm s}$ of the bilayer model for several interaction strengths.
From this figure,
we can see the same tendency as in the single-orbital case.
Namely,
$\alpha_{\rm s}$ increases with lowering the temperature or with increasing the interaction strength.
The difference between the one-shot and self-consistent calculations is monotonic in terms of the interaction strength as in the single-orbital case.
However, 
the difference is smaller than that in the single-orbital case.
This is because the interaction strength effectively becomes smaller than that of the single-orbital case 
due to the existence of the interlayer hopping $t_{\perp}$.

Figure~\ref{fig:2021-03-19-12-11}~(b) shows the spectral functions obtained from IPT + parquet and IPT + parquet + EDF~[self-consistent calculation] 
at the temperature where the spin fluctuation is large~($\alpha_{\rm s} \sim 1$).
We can see the pseudo gap structure in all interaction strengths.
In comparison with the single-orbital case,
the depth of dips between the central Kondo peak and Hubbard bands are small,
so we can see that the interaction strengths are effectively small as mentioned above.

\section{Results of S2F + EDF}\label{sec:2020-12-25-04-06}

In this section, 
we show the results obtained from the calculations for the nonlocal correlation using the combination of S2F and EDF methods.
Here,
we compare three calculation procedures:
IPT + parquet + EDF, IPT + parquet + S2F + EDF, and CT-QMC + S2F + EDF~[we also call these three procedures by abbreviations IE, ISE, and CSE, respectively].
In IPT + parquet + S2F + EDF, we solve the impurity problem by IPT + parquet. 
After that, we estimate the full vertex from the self-energy by S2F.
Using this full vertex, we take into account the nonlocal correlation into DMFT solution by EDF.
CT-QMC + S2F + EDF and IPT + parquet + S2F + EDF are the same except for the impurity solvers.
In CT-QMC + S2F + EDF, we solve the impurity problem by CT-QMC.
Results of IPT + parquet + EDF, which has already been shown in Sec.~\ref{sec:2020-12-24-20-47},
are once again shown here for comparison.
Also, 
the results shown here are obtained by the one-shot calculation in terms of EDF, 
i.e.,
we perform only one EDF calculation after solving the impurity problem.
Hence, no self-consistency is imposed.

\begin{figure}[] 
  \centering
  {\includegraphics[width=70mm,clip]{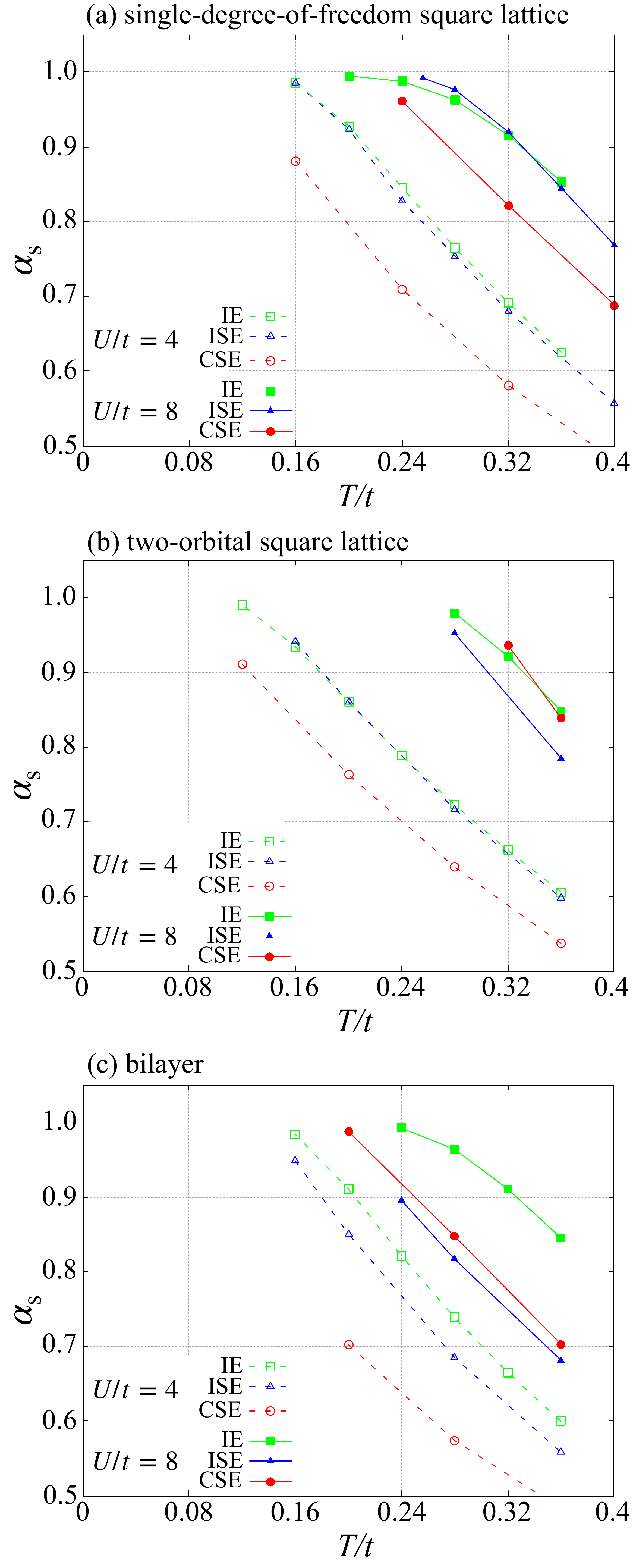}} 
  \caption{
    The temperature dependence of $\alpha_{\rm s}$ of 
    (a) the single-orbital square lattice model,
    (b) the two-orbital square lattice model,
    and 
    (c) the bilayer model.
    Green, blue, and red lines indicate the results obtained by IPT + parquet + EDF~(IE), IPT + parquet + S2F + EDF~(ISE), and CT-QMC + S2F + EDF~(CSE), respectively.
    Open symbols indicate the results at $U/t=4$ and closed symbols at $U/t=8$.
  } 
  \label{fig:2020-12-24-20-11}
\end{figure}

Figure~\ref{fig:2020-12-24-20-11} shows the temperature dependence of $\alpha_{\rm s}$ of three models:
the single-orbital square lattice, the two-orbital square lattice, and the bilayer.
First,
we compare the results of IPT + parquet + EDF~(IE) and IPT + parquet + S2F + EDF~(ISE), 
which are different only in the process taking into account the spatial correlation.
From Fig.~\ref{fig:2020-12-24-20-11}~(a), 
we can see very good agreement with these two procedures in the single-orbital square lattice model. 
In the two-orbital square lattice model~[Fig.~\ref{fig:2020-12-24-20-11}~(b)], 
we can see the difference between the IE and the ISE at $U/t=8$
while 
the difference is hardly seen at $U/t=4$.
In the bilayer model~[Fig.~\ref{fig:2020-12-24-20-11}~(c)], 
the difference between the two procedures can be seen at both interaction strengths,
and especially at $U/t=8$, 
ISE and  IE are largely different.
Note that, 
at present, 
we can not tell which of IE and ISE is better.
Comparison with some numerically exact method to judge this point remains as future work.
Next,
we compare the results of CT-QMC + S2F + EDF~(CSE) and IPT + parquet + S2F + EDF~(ISE),
which are different only in the impurity solver.
From Fig.~\ref{fig:2020-12-24-20-11},
we can see the difference between the two procedures in all three models.
$\alpha_{\rm s}$'s obtained from CSE are smaller than that of ISE except at $U/t=8$ in the two-orbital square lattice and the bilayer models.
These differences can be understood from the difference in the results of the solution of the (local) impurity problems shown in Ref.~\cite{PhysRevB.104.035160}.
Except at $U/t=8$ in the two-orbital square lattice and bilayer models, 
the correlation effects, which can be seen in the quasi-particle weight $Z$ for example, 
are smaller in CT-QMC than in IPT + parquet.
Then, 
the full vertex which acts as the bare vertex in dual fermion system is smaller in CT-QMC than in IPT + parquet.
Therefore, 
$\alpha_{\rm s}$'s in CSE are smaller than that in ISE.
At $U/t=8$ in the two-orbital square lattice and the bilayer models, 
the above relation becomes the opposite.

\section{Discussion}\label{sec:2021-04-17-01-37}

\subsection{Computational efficiency}
Here, we discuss the computational efficiency of the methods developed in this study.
Before we move on,
we recall the difference between two types of extensions for spatial fluctuation, the cluster type and the diagrammatic type.
In the cluster extensions, 
we do not have to calculate two-particle quantities, so that solving the impurity problem is not so difficult, 
whereas the system size is quite restricted.
The cluster size is about $4-16$~\cite{PhysRevLett.86.139,PhysRevB.80.245102,PhysRevB.82.155101}. % in many cases.
On the other hand, in the diagrammatic extensions, 
we can take large system sizes such as $32\times 32$, $64 \times 64$, or more $k$-meshes~\cite{PhysRevB.90.235132,PhysRevB.97.115150,PhysRevB.98.155117}, 
but we have to calculate the two-particle quantities. 
Calculating the two-particle quantities with exact impurity solvers, especially in multiband systems, is challenging.
Furthermore,
the reduction of the number of Matsubara frequencies is indispensable to perform the calculation in a practical computational time, so that calculations at low temperatures are limited.
By using S2F, in any solver,
we can  overcome the difficulty in estimating the two-particle quantities.
In EDF, 
we can take large system sizes and the reduction of  the number of Matsubara frequencies is not necessarily required, 
due to its low numerical cost.
Moreover, 
in IPT + parquet + EDF, which is a combination of IPT + parquet~\cite{PhysRevB.104.035160} and EDF,
the local and nonlocal calculations are both numerically efficient.

We can roughly estimate the core hours
\footnote{
  (core hours) = (the number of CPU cores we use) $\times$ (the number of hours for a calculation)
}
of CT-QMC + D${\rm \Gamma}$A~(diagrammatic type) in two-band cases from the information of the previous study~\cite{doi:10.7566/JPSJ.87.041004}.
The core hours of CT-QMC + D${\rm \Gamma}$A is $ {\cal O}(10^{4})$, while IPT + parquet + EDF ${\cal O}(1)$. 
Furthermore, 
the core hour of CT-QMC + D${\rm \Gamma}$A increases more rapidly than that of IPT + parquet + EDF with increasing the number of Matsubara frequencies.
Hence, the CT-QMC + D${\rm \Gamma}$A calculation is limited 
when the temperature is low or 
the number of bands 
of the system is large. 
The cost of the standard dual fermion (not EDF) is assumed to be the same as D${\rm \Gamma}$A since the procedures of these methods are similar.
The comparison of the computational efficiency between the previous methods and the newly developed methods is given in Table~\ref{fig:2020-12-24-22-21}.
Finally, 
we show the computational scales of our methods.
The most expensive part of IPT + parquet and S2F scales as $(N_{\omega} \log N_{\omega}) N_{b}^{4}$ 
and 
EDF as $N_{\omega}N_{k}\log (N_{\omega}N_{k})N_{b}^{4}$,
where $N_{\omega}$, $N_{k}$, and $N_{b}$ are the numbers of Matsubara frequencies, $k$-meshes, and bands, respectively.
These estimations assume the conditions adopted in this study, in which the process of the Fourier transformation is most expensive.
If $N_{b}$ increases and 
the multiplication of two-particle quantities becomes the most expensive part, 
the scales of IPT + parquet and S2F become $N_{\omega}N_{b}^{6}$ and EDF $N_{\omega}N_{k}N_{b}^{6}$.

\begin{table*}[]
  \caption{
    Comparison of computational efficiency.
  %  ``EIS'' means exact impurity solvers.
  %  Blue circle(red cross) indicates advantage(disadvantage) of each method.
  }
  \centering
  {\includegraphics[width=150mm,clip]{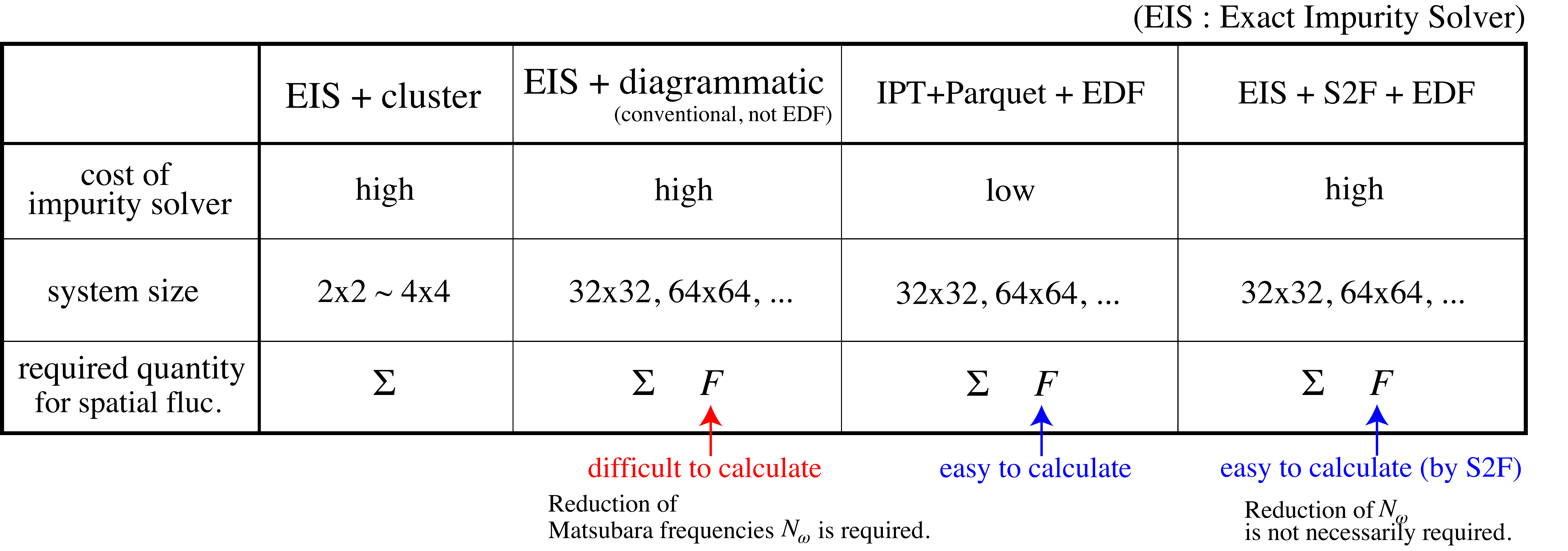}} 
  \label{fig:2020-12-24-22-21}
\end{table*}

\subsection{Connection to experiments}
By using the methods developed here and Ref.~\cite{PhysRevB.104.035160}, 
we can obtain the two-particle quantities such as the spin susceptibility $\chi_{\rm s}(\bm{q},\nu)$ as well as the one-particle quantities such as the spectral function $A(\bm{k},\omega)$ with high spatial resolution. 
The spectral function $A(\bm{k},\omega)$ corresponds to the data obtained in angle-resolved photoemission spectroscopy~(ARPES) 
and the susceptibilities $\chi_{\rm c,s}(\bm{q},\nu)$ corresponds to the data obtained in inelastic neutron scattering or the nuclear magnetic resonance~(NMR). 
Combining the present methods with {\it ab initio} methods, which is a future work, enables us to compare
theoretical results and experimental results or to explain experimental results,
even in the multiband and strongly correlated systems in which we cannot carry out calculations by conventional methods.
In addition, our methods enable us to search for good conditions of physical properties such as high-temperature superconductivity in multiband and strongly correlated systems.
In conventional methods,  it is impossible to analyze the vast parameter~(the filling, difference of energy level, correlation strength,..) space comprehensively even in systems with relatively small degrees of freedom.
The methods developed in this study and Ref.~\cite{PhysRevB.104.035160} enable us to do this.
The development of our understanding of the relation between these parameters and physical properties may lead to suggestions of novel materials.

\subsection{Possible improvements}\label{sec:2021-08-03-16-22}

\subsubsection{Improvement of S2F}\label{sec:2021-08-10-13-34}
As mentioned in Sec.~\ref{sec:2021-04-17-01-36}, 
in S2F, 
the diagonal structures of ph and ${\rm \overline{ph}}$ channels of the full vertex, 
namely, 
the vertices $\Phi_{c}$ and $ \Phi_{s}$ tend to be underestimated.
Let us consider a possible improvement of S2F. 
When being faithful to the parquet formalism, 
the irreducible vertices $\Gamma_{r}$ ($r=$ph,${\rm \overline{ph}}$,pp) have to contain the contribution of the cross and central structures,
but they do not in the present formalism of S2F.
Hence, a possible improvement of S2F is to take into account these contributions.
To be specific, 
we determine the correction factor $C$ in the self-consistent procedure which involves an extended version of simplified parquet method, 
in which the irreducible vertices $\Gamma_{r}$ ($r=$ph,${\rm \overline{ph}}$,pp) also contain the contribution of the cross and central structures.
To give $\Gamma_{r}$ these contributions,
we modify Eq.~(\ref{eq:2020-05-10-21-21})  in Appendix.~\ref{sec:2020-10-09-00-10} as 
\begin{align}
  &\tilde{\Lambda}_{r,\alpha\beta\gamma\lambda}(\omega_{n},\omega_{n'},\nu_{m}) \nonumber \\
  & \hspace{10pt}= z_{r} C_{\alpha\alpha'}(\omega_{n})C_{\beta\beta'}(\omega_{n}+\nu_{m}){\Lambda}_{r,\alpha'\beta'\gamma'\lambda'} C_{\gamma'\gamma}(\omega_{n'})C_{\lambda'\lambda}(\omega_{n'}+\nu_{m}).
  \label{eq:2021-07-30-20-40}
\end{align}
However, in the practical calculation, 
we have to make the Green's function-like and susceptibility-like functions by moving the correction factor from the vertices $\tilde{\Lambda}$ as in EDF in Sec.~\ref{sec:2020-07-10-21-41}: 
\begin{align}
 % \tilde{G}^{C_{i}}(k) =& C_{i}(\omega_{n})G_{\rm dual}(k) 
 % \label{eq:2020-06-29-15-25} \\
  {G}^{C}(k) =& C(\omega_{n})G(k) C(\omega_{n}),
  \label{eq:2021-07-30-15-26} \\
  {\chi}_{0}^{C}(q) =& -\sum_{k} {G}^{C}(k){G}^{C}(k+q),
  \label{eq:2021-07-30-15-27}
\end{align}
and 
perform the simplified parquet calculation with these functions
to keep the numerical cost low.
Hence,
in the practical calculation,
we do not modify  Eq.~(\ref{eq:2020-05-10-21-21}) but 
replace the Green's function and the susceptibility with the quantities in Eqs.~(\ref{eq:2021-07-30-15-26}) and  (\ref{eq:2021-07-30-15-27}).

The S2F procedure given in Sec.~\ref{sec:2020-07-10-21-40} is modified as follows.
\begin{enumerate}
  \item[(i)] calculate $F_{0}$ by the extended version of simplified parquet method,
    in which we replace the Green's function and the susceptibility with the quantities in Eqs.~(\ref{eq:2021-07-30-15-26}) and  (\ref{eq:2021-07-30-15-27}) in the procedure shown in Appendix.~\ref{sec:2020-10-09-00-10}.  
  \item[(ii)] calculate $X$ by Eq.~(\ref{eq:2020-06-29-13-05}).  
  \item[(iii)] obtain $C$ by  $C=\Sigma^{\rm CR}X^{-1}$. 
  \item[(iv)] Go back to (i) (iterate until convergence). 
\end{enumerate}
With this modification, 
we expect that the agreement between the results of S2F and the exact methods~(CT-QMC, ED) can be improved.

\subsubsection{Improvements of EDF} \label{sec:2021-08-10-13-35}
As mentioned in Sec.~\ref{sec:2021-03-19-14-19}, 
EDF tends to overestimate the spin fluctuation.
There can be three reasons:
(i) the omission of  the local vertices in ${\rm \overline{ph}}$ and pp channels in Eq.~(\ref{eq:2020-06-29-14-59}), 
(ii) the adoption of the ladder approximation in the dual fermion system, 
(iii) the overestimation of the cross and central structures given by $C$ (although we are not certain about this at present.)
Let us consider possible improvements for (i) and (ii) in the following,
since the overestimation of the cross and the central structures given by $C$ is expected to be suppressed by the modification mentioned in the previous section~(Sec.~\ref{sec:2021-08-10-13-34}).

(i)~A way to take into account the contribution of ${\rm \overline{ph}}$ and pp channels omitted in  Eq.~(\ref{eq:2020-06-29-14-59}) is to replace the vertex $\Lambda$ with $z\Lambda$, where $z$ is the constant renormalization factor in the simplified parquet method~(see Appendix.~\ref{sec:2020-10-09-00-10}).
$z$ contains the contribution of ${\rm \overline{ph}}$ and pp channels, so the overestimation of the spin fluctuation can be suppressed.

(ii)~We can adopt the simplified parquet method also in the dual fermion system instead of the ladder approximation. 
We can consider the competition of the fluctuations originating from multiple channels by using the parquet equations, 
and by doing so,
the overestimation of the spin fluctuation is expected to be suppressed.
Indeed, 
the antiferromagnetic and the superconducting states in the phase diagram of the cuprate superconductor were well reproduced by solving the dual fermion system with the parquet equations in a previous study~\cite{PhysRevB.101.075109}.   
In Ref.~\cite{PhysRevB.101.075109}, 
authors performed the dual fermion calculation  with only the lowest Matsubara frequency using  the effective low-frequency  model, 
which was constructed by integrating out the high-frequency part of the dual fermion variables in the dual action. 
On the contrary, they adopted the full parquet equations in terms of the momentum domain. 
Hence, the numerical cost was very high even though they used only the lowest Matsubara frequency.
If we adopt the simplified parquet method in the dual fermion system, 
we can treat the competing fluctuations in terms of not only the momentum but also the frequency domain with sufficiently low numerical cost,
although the treatment of the momentum domain can be less accurate.

\section{Conclusion}\label{sec:2021-04-17-01-38}

We have shown that the local full vertex can be approximated in a simple form.
By using the simplified form of the local full vertex,
we have developed two methods for the nonlocal fluctuation.
One is to estimate the two-particle full vertex  from the one-particle self-energy~(S2F).
This enables us to combine the extensions for the spatial fluctuation with any impurity solver.
The other is the efficient calculation method in the dual fermion method~(EDF).
In S2F with CT-QMC,
we have shown that the local full vertex obtained by S2F procedure is qualitatively consistent with that obtained by the exact methods: CT-QMC and ED.
In IPT + parquet + EDF, 
we could have seen the temperature dependence of $\alpha_{\rm s}$ and the pseudo gap structure in the spectral function which are consistent with the previous studies in the single-orbital system,
and
have found the similar tendencies and structures in the multiband systems.
%and the competition between the formation of Kondo singlet and the spatial spin fluctuation.
%Especially, the latter can not be seen in the one-shot calculation and so this shows the importance of the self-consistent calculation. 
In the calculations with S2F + EDF procedure,
we have recognized that there are differences between the results of IPT + parquet + EDF and IPT + parquet + S2F + EDF in the multiband systems.
At present, we can not tell which is better.
We need to compare the results with some numerically exact methods to judge this point, which is future work.
Also, we have confirmed that, using the S2F procedure,  we can take into account the spatial fluctuation into the CT-QMC results without directly calculating the two-particle quantities.
In addition, numerical costs are largely reduced.
%Therefore,
We expect that our methods developed in this study can be useful for analyzing various %multiband and 
strongly correlated systems.

\begin{acknowledgements}
  Part of the numerical calculations was performed using the large-scale computer systems provided by the following institutions: 
  the supercomputer center of the Institute for Solid State Physics, the University of Tokyo, 
  and 
  the Information Technology Center, the University of Tokyo.
  This study has been supported by JSPS KAKENHI Grants No.JP18H01860.
\end{acknowledgements}

\appendix

\section{Simplified parquet method}\label{sec:2020-10-09-00-10}
In this section,
we introduce the simplified parquet method developed in Ref.~\cite{doi:10.1143/JPSJ.79.094707} and extended for multiband systems in Ref.~\cite{PhysRevB.104.035160}, 
in which the numerical cost is much lower than that of the non-simplified parquet method since 
we should practically consider just one of the three variables $(k,k',q)$.
Before we start introducing the simplified parquet method,
we define the following notation which indicates the set of the degrees of freedom, the frequencies, and wave vectors.
\begin{align}
  D =& (\alpha,\beta,\gamma,\lambda), (k,k',q) \label{eq:2020-05-10-14-49} \\
  T =& (\alpha,\beta,\lambda,\gamma), (k,-q-k',q) \label{eq:2021-08-12-14-54} \\ 
  C =& (\alpha,\gamma,\beta,\lambda), (k,k+q,k'-k) \label{eq:2020-05-10-14-51} \\
  P =& (\alpha,\lambda,\gamma,\beta), (k,k',-q-k-k') \label{eq:2020-05-10-14-52} \\
  X =& (\alpha,\gamma,\lambda,\beta), (k,-k-q,k'-k) \label{eq:2020-05-10-14-53}
\end{align}

In the presence of SU(2) symmetry in spin space, 
the full vertex can be divided into four channels 
$c$(charge), $s$(spin), $e$(even), $o$(odd)
in terms of the parity of spin.
\begin{align}
  F_{r}(D) =& \Lambda_{r}(D) + \Phi_{ {\rm ph}, r}(D) + \Phi_{ {\rm \overline{ph}},r}(D) + \Phi_{ {\rm pp},r}(D) \hspace{10pt} (r = {c,s,e,o} )
  \label{eq:2020-06-14-15-03}
\end{align} 
We can rewrite 
Eq.~(\ref{eq:2020-06-14-15-03})
as follows by replacements of variables and indices.
\begin{align}
  F_{c}(D) =& \Lambda_{c}(D) + \Phi_{{\rm ph},c}(D) \nonumber \\ &\hspace{-20pt}- \dfrac{1}{2}[ \Phi_{{\rm ph},c} + 3\Phi_{{\rm ph},s} ](C)  + [ \Phi_{{\rm pp},e} - 3\Phi_{{\rm pp}, o} ](P) \label{eq:2020-05-12-13-58} \\  
  F_{s}(D) =& \Lambda_{s}(D) + \Phi_{{\rm ph},s}(D) \nonumber \\ &\hspace{-20pt}- \dfrac{1}{2}[ \Phi_{{\rm ph},c} -  \Phi_{{\rm ph},s} ](C)  - [ \Phi_{{\rm pp},e} -  \Phi_{{\rm pp}, o} ](P) \label{eq:2020-05-12-13-59} \\  
  F_{e}(D) =& \Lambda_{e}(D) + \Phi_{{\rm pp},e}(D) \nonumber \\ &\hspace{-20pt}+ \dfrac{1}{4}[ \Phi_{{\rm ph},c} - 3\Phi_{{\rm ph},s} ](X)  + \dfrac{1}{4}[ \Phi_{{\rm ph},c} - 3\Phi_{{\rm ph}, s} ](P) \label{eq:2020-05-12-14-00} \\  
  F_{o}(D) =& \Lambda_{o}(D) + \Phi_{{\rm pp},o}(D) \nonumber \\ &\hspace{-20pt}+ \dfrac{1}{4}[ \Phi_{{\rm ph},c} +  \Phi_{{\rm ph},s} ](X)  - \dfrac{1}{4}[ \Phi_{{\rm ph},c} + \Phi_{{\rm ph}, s} ](P) \label{eq:2020-05-12-14-01}   
\end{align} 
As we can see from 
Eqs.~(\ref{eq:2020-05-12-13-58})$-$(\ref{eq:2020-05-12-14-01}),
since $c,s$ always appear together with ph, 
$e,o$ with pp, 
we omit the subscripts ph or pp hereafter. 
We write the third and fourth term as $\gamma^{(1)}_{r}$ and $\gamma^{(2)}_{r}$, respectively.
To say, 
\begin{align}
  \hat{\gamma}^{(1)}_{c} =& - \dfrac{1}{2}[ \hat{\Phi}_{c} + 3\hat{\Phi}_{s} ],  \hspace{10pt} \hat{\gamma}^{(2)}_{c} =                [ \hat{\Phi}_{e} - 3\hat{\Phi}_{o} ]      \label{eq:2021-04-27-01-02} \\ 
  \hat{\gamma}^{(1)}_{s} =& - \dfrac{1}{2}[ \hat{\Phi}_{c} -  \hat{\Phi}_{s} ],  \hspace{10pt} \hat{\gamma}^{(2)}_{s} =  -             [ \hat{\Phi}_{e} -  \hat{\Phi}_{o} ]    \label{eq:2021-04-27-01-03} \\
  \hat{\gamma}^{(1)}_{e} =&   \dfrac{1}{4}[ \hat{\Phi}_{c} - 3\hat{\Phi}_{s} ],  \hspace{10pt} \hat{\gamma}^{(2)}_{e} =    \dfrac{1}{4}[ \hat{\Phi}_{c} - 3\hat{\Phi}_{s} ] \label{eq:2021-04-27-01-04} \\
  \hat{\gamma}^{(1)}_{o} =&   \dfrac{1}{4}[ \hat{\Phi}_{c} +  \hat{\Phi}_{s} ],  \hspace{10pt} \hat{\gamma}^{(2)}_{o} =  - \dfrac{1}{4}[ \hat{\Phi}_{c} +  \hat{\Phi}_{s} ] \label{eq:2021-04-27-01-05} 
\end{align}
We can also write the 
Bethe-Salpeter equation by using the four channels: 
\begin{align}
  \hat{F}_{r} &= \hat{\Gamma}_{r} + \hat{\Phi}_{r} \hspace{10pt} (r = {c,s,e,o}) \label{eq:2020-05-10-13-33} \\
  \hat{\Phi}_{r} &= -\hat{\Gamma}_{r}\hat{\chi}_{0} \hat{F}_{r} = -\hat{\Gamma}_{r}\hat{\chi}_{r}\hat{\Gamma}_{r},
  \label{eq:2020-05-10-13-34} 
\end{align}
and the susceptibilities: 
\begin{align} 
  \hat{\chi}_{r} =& \hat{\chi}_{0} - \hat{\chi}_{0}\hat{\Gamma}_{r}\hat{\chi}_{r} = \hat{\chi}_{0} - \hat{\chi}_{0} \hat{F}_{r}\hat{\chi}_{0}.
  \label{eq:2020-05-12-14-57}
\end{align}

With this preliminary, we will explain the details of the approximation in the simplified parquet method.
First, 
we use the bare vertices $U_{r}$ as the fully irreducible vertices $\Lambda_{r}$:
\begin{align}
  \Lambda_{c}(D) =&                U_{c}(D) \label{eq:2020-05-12-15-37} \\
  \Lambda_{s}(D) =&               -U_{s}(D) \label{eq:2020-05-12-15-38} \\
  \Lambda_{e}(D) =&   \dfrac{1}{4}(U_{c} + 3U_{s} )(P) \label{eq:2020-05-12-15-39} \\
  \Lambda_{o}(D) =&  -\dfrac{1}{4}(U_{c} -  U_{s} )(P) \label{eq:2020-05-12-15-40}
\end{align}
We calculate the susceptibilities by using the random phase approximation~(RPA) type formula:
\begin{align} 
  \hat{\chi}_{r}(q) =& \hat{\chi}_{0}(q)[\hat{I} + \hat{\tilde{\Lambda}}_{r}\hat{\chi}_{0}(q)]^{-1}.
  \label{eq:2020-05-10-21-27} 
\end{align}
where 
\begin {align}
  \hat{\tilde{\Lambda}}_{r} =& z_{r} \hat{\Lambda}_{r} 
  \label{eq:2020-05-10-21-21}
\end{align}
and $z_{r}$ is the constant renormalization factor.
With these, 
the irreducible vertices can be calculated as 
\begin{align} 
  \hat{\Phi}_{r} =&
  -\hat{\tilde{\Lambda}}_{r} \hat{\chi}_{r} \hat{\tilde{\Lambda}}_{r} .
  \label{eq:2020-05-12-16-20} 
\end{align}
By this approximation,
the generalized momentum dependences in Eqs.~(\ref{eq:2020-05-10-14-49})-(\ref{eq:2020-05-10-14-53}) are replaced as
\begin{align}
  D :&\ (k,k',q) \to q \label{eq:2021-04-22-02-08} \\
  C :&\ (k,k+q,k'-k) \to k'-k \label{eq:2021-04-22-02-09} \\
  P :&\ (k,k',-q-k-k') \to -q-k-k'\label{eq:2021-04-22-02-10} \\
  X :&\ (k,-k-q,k'-k) \to k'-k \label{eq:2021-04-22-02-11}
\end{align}
If we consider the local case,
Eqs.~(\ref{eq:2021-04-22-02-08})-(\ref{eq:2021-04-22-02-11}) mean that the full vertex has only the diagonal structure.

From the comparison between
susceptibilities from the RPA type Eq.~(\ref{eq:2020-05-10-21-27}) and the parquet type Eq.~(\ref{eq:2020-05-12-14-57}),
we can obtain the renormalization factor $z_{r}$ as 
\begin{align}
  z_{r} =&
  1 + 
  \dfrac{ {\rm Tr} \bigl[ \hat{\chi}_{0}(k,q)( \hat{\gamma}^{(1)}_{r}(k-k') + \hat{\gamma}^{(2)}_{r}(k+k'+q) ) \hat{\chi}_{0}(k',q) \bigr]}{ {\rm Tr} \bigl[\hat{\chi}_{0}(q)\hat{\Lambda}_{r}\hat{\chi}_{0}(q)\bigr]},%  \hspace{10pt} 
  \label{eq:2020-05-10-16-29}
\end{align}
where ${\rm Tr}A = \sum_{k,k',q}\sum_{\alpha} A_{\alpha \alpha \alpha \alpha}(k,k',q)$. 
Although the summation in the numerator of Eq.~(\ref{eq:2020-05-10-16-29}) is taken over $k,k',q$, 
we can rewrite it as a summation over $q$ by a variable conversion.
Hence, 
we treat only $q$ practically.
%Repeating these procedure, we determine $z_{r}$ self-consistently.
The calculation procedure of the simplified parquet method is as follows.
\begin{enumerate}
  \item
    Calculate the bare vertices $\Lambda_{r}$  by Eqs.~(\ref{eq:2020-05-12-15-37})-(\ref{eq:2020-05-12-15-40}).
  \item
    Calculate the renormalized vertices $\tilde{\Lambda}_{r}$ by Eq.~(\ref{eq:2020-05-10-21-21}). \\ 
    The initial values are $(z_{c},z_{s},z_{e},z_{o})=(1,0.1,1,1)$.  
  \item
    Calculate the susceptibilities $\chi_{r}$ by Eq.~(\ref{eq:2020-05-10-21-27}).
  \item
    Calculate the reducible vertices $\Phi_{r}$ by Eq.~(\ref{eq:2020-05-12-16-20}).
  \item
    Calculate the vertices $\gamma^{(1)}_{r}$ and $\gamma^{(2)}_{r}$ by  Eqs.~(\ref{eq:2021-04-27-01-02})-(\ref{eq:2021-04-27-01-05}).
  \item
    Update the renormalization factor $z_{r}$ by Eq.(\ref{eq:2020-05-10-16-29}).
  \item
    Go back to step 2. (until convergence).
\end{enumerate}
After convergence, 
we already have obtained the vertices $\Phi_{r}$, $F_{r}$ and the susceptibilities $\chi_{r}$.

If we obtain the full vertex by the above procedure, we can obtain the self-energy as follows.
\begin{align}
  \Sigma_{\alpha\beta}(k)
  =&
  \dfrac{1}{4}\sum_{\gamma\lambda}\sum_{q} \Bigl[  \hat{F}_{c}(q) \hat{\chi}_{0}(q)\hat{U}_{c} + 3\hat{F}_{s}(q)\hat{\chi}_{0}(q)\hat{U}_{s} \Bigr]_{\alpha\gamma\beta\lambda} G_{\gamma\lambda}(k+q) 
  \label{eq:2020-06-28-17-38}
\end{align}
In practical calculation, however, we omit the contribution from pp channel in the self-energy since it tends to be overestimated.

\

\section{Dual fermion method}\label{sec:2021-01-01-01-30}
The dual fermion method 
\cite{PhysRevB.77.033101,PhysRevB.79.045133, PhysRevB.90.235132,PhysRevB.97.115150,PhysRevB.98.155117}
is one of the extensions of DMFT to take into account the spatial fluctuation,
which is accomplished by introducing an auxiliary particle called dual fermion.
In the dual fermion system,
particles interact with each other by the interaction which includes the local correlation effects of the original lattice system.
Although there are some differences,
we can adopt the method of the diagram expansion as in the lattice systems.

\subsection{Outline}

The effective action of the Hubbard model expressed by Grassmann variables is  
\begin{align}
  S[c,c^{*}] 
  =&
  -\sum_{n \bm{k}\sigma}c^{*}_{\omega\bm{k}\sigma} \Bigl(i\omega_{n} + \mu - \epsilon_{\bm{k}}\Bigr)c_{\omega\bm{k}\sigma}
  + U \sum_{i}\int d\tau \hspace{2pt} n_{\tau i\uparrow} n_{\tau i\downarrow} ,
  \label{eq:2020-05-30-18-11}
\end{align}
and the Anderson model 
\begin{align} 
  S_{\rm imp}[c_{i},c_{i}^{*}] 
  =&
  -\sum_{n\sigma} c^{*}_{\omega i\sigma}\Bigl(i\omega_{n}+\mu-\Delta(i\omega_{n}) \Bigr) c_{\omega i\sigma}
  + U \int d\tau \hspace{2pt} n_{\tau i\uparrow} n_{\tau i\downarrow} .
  \label{eq:2020-05-30-18-27}
\end{align}
%$c_{i},c^{*}_{i}$ in Eq.~(\ref{eq:2020-05-30-18-27}) corresponds to 
%$f,f^{*}$ in Eq.~(\ref{eq:2020-05-28-03-08})
%due to the equivalency in DMFT.
From Eqs.~(\ref{eq:2020-05-30-18-11}) and (\ref{eq:2020-05-30-18-27}), 
we can obtain 
\begin{align} 
  S[c,c^{*}] 
  =&
  \sum_{i}S_{\rm imp}[c_{i},c^{*}_{i}]
  +
  \sum_{n\bm{k}\sigma} c^{*}_{\omega \bm{k}\sigma} \Bigl( \Delta(i\omega_{n}) - \epsilon_{\bm{k}} \Bigr)c_{\omega\bm{k}\sigma} .
  \label{eq:2020-05-30-18-35}
\end{align}
Here, 
we introduce new Grassmann variables $d,d^{*}$ 
and 
use the following identity 
\begin{align}
  \exp&\Bigl( A^{2} c^{*}_{\omega\bm{k}\sigma}c_{\omega\bm{k}\sigma} \Bigr) \nonumber \\
  =&
  B^{-2} \int {\cal D}d^{*}{\cal D}d
  \exp \Bigl[
    -AB \Bigl( c^{*}_{\omega\bm{k}\sigma}d_{\omega\bm{k}\sigma} + d^{*}_{\omega\bm{k}\sigma} c_{\omega\bm{k}\sigma} \Bigr) 
    \nonumber \\
    &\hspace{60pt}-
    B^{2}d^{*}_{\omega\bm{k}\sigma}d_{\omega\bm{k}\sigma}
  \Bigr],
  \label{eq:2020-05-30-18-43}
\end{align}
where 
$A,B$ are complex numbers. 
Assuming 
$A^{2}=\Delta(i\omega_{n}) - \epsilon_{\bm{k}}, \hspace{3pt} B^{-2}=G_{\rm imp}^{-2}(i\omega_{n})(\Delta(i\omega_{n})-\epsilon_{\bm{k}})^{-1}$,
the partition function in the lattice system can be transformed as 
\begin{align}
  Z 
  =&
  Z_{d}\int {\cal D}d^{*}{\cal D}d{\cal D}c^{*}{\cal D}c \hspace{3pt} \exp( - S[c,c^{*},d,d^{*}] ),
  \label{eq:2020-05-30-18-52} 
\end{align}
\begin{align}
  S[c,c^{*},d,d^{*}] 
  =&
  \sum_{i}S_{\rm imp}[c_{i},c^{*}_{i}] \nonumber \\
  +& 
  \sum_{n\bm{k}\sigma} \Bigl[
    G_{\rm imp}^{-1}(i\omega_{n}) \Bigl( d^{*}_{\omega\bm{k}\sigma} c_{\omega\bm{k}\sigma} + c^{*}_{\omega\bm{k}\sigma} d_{\omega\bm{k}\sigma} \Bigr) \nonumber \\
    & \hspace{0pt}  +
    G_{\rm imp}^{-2}(i\omega_{n}) \Bigl(\Delta(i\omega_{n})-\epsilon_{\bm{k}}\Bigr)^{-1} d^{*}_{\omega\bm{k}\sigma}d_{\omega\bm{k}\sigma}
  \Bigr],
  \label{eq:2020-05-30-18-57}
\end{align}
where 
$Z_{d}=\prod_{n\bm{k}} G^{2}_{\rm imp}(\omega_{n}) (\Delta(i\omega_{n})-\epsilon_{\bm{k}})$.
The second term in the right hand side of 
Eq.~(\ref{eq:2020-05-30-18-57})
can be transformed to the real space representation 
since $G_{\rm imp}(i\omega_{n})$ is independent of $\bm{k}$. 
Therefore $S$ is rewritten as 
\begin{align} 
  S[c,c^{*},d,d^{*}] 
  =&
  \sum_{i} S_{\rm site}[c_{i},c^{*}_{i},d_{i},d^{*}_{i}] \nonumber\\ 
  +& \sum_{n\bm{k}\sigma} G_{\rm imp}^{-2}(i\omega_{n})\Bigl(\Delta(i\omega_{n})-\epsilon_{\bm{k}}\Bigr)^{-1} d_{\omega\bm{k}\sigma}^{*}d_{\omega\bm{k}\sigma},
  \label{eq:2020-05-31-00-02}
\end{align}
where
\begin{align}
  S_{\rm site}[c_{i},c^{*}_{i},d_{i},d^{*}_{i}] 
  =&
  S_{\rm imp}[c_{i},c^{*}_{i}] \nonumber \\
  +& \sum_{n i\sigma}G^{-1}_{\rm imp}(i\omega_{n}) (d^{*}_{\omega i \sigma} c_{\omega i\sigma} + c^{*}_{\omega i\sigma}d_{\omega i\sigma} ).
  \label{eq:2020-05-31-00-03}
\end{align}
The crucial point here is that 
the integration over the initial variables 
$c_{i},c^{*}_{i}$
can be performed separately for each site.
Executing the integration after the Taylor expansion in terms of $d,d^{*}$,
we obtain 
\begin{align}
  \int& {\cal D}c^{*}_{i} {\cal D}c_{i} \hspace{2pt} \exp(-S_{\rm site}) \nonumber \\
  &=
  Z_{\rm imp} \exp 
  \Bigl( 
    \sum_{n\sigma}G^{-1}_{\rm imp}(i\omega_{n})d^{*}_{\omega i \sigma}d_{\omega i \sigma} \nonumber \\
    &-
    \dfrac{1}{4}\sum_{nn'm}F_{\rm imp}(i\omega_{n},i\omega_{n'},i\nu_{m})d^{*}_{\omega i \sigma} d_{\omega+\nu i \sigma} d_{\omega' i \sigma'} d^{*}_{\omega'+\nu i \sigma'} 
    \nonumber \\
    &\hspace{90pt}+
    \cdots
  \Bigr),
  \label{eq:2020-05-31-01-33}
\end{align}
where 
$F_{\rm imp}$ is the local full vertex derived by solving the impurity problem.
Therefore 
the effective action in the dual fermion system is given by 
\begin{align}
  S[d,d^{*}]
  =&
  -\sum_{n\bm{k}\sigma} 
  {G}_{\rm 0dual}^{-1}(k) 
  d^{*}_{\omega\bm{k}\sigma}d_{\omega\bm{k}\sigma}
  +
  \sum_{i} V[d_{i},d^{*}_{i}],
  \label{eq:2020-05-31-02-37} \\
  {G}_{\rm 0dual}(k)
  =&
  \Bigl( i\omega_{n} + \mu  - \epsilon_{\bm{k}} - \Sigma(i\omega_{n}) \Bigr)^{-1} - G_{\rm imp}(i\omega_{n}),
  \label{eq:2020-05-31-02-48} 
%  V[d_{i},d^{*}_{i}]
%  =&
%  \dfrac{1}{4}F_{\rm imp}(i\omega_{n},i\omega_{n'},\nu_{m})d^{*}_{\omega i \sigma} d_{\omega+\nu i \sigma} d_{\omega' i \sigma'} d^{*}_{\omega'+\nu i \sigma'} 
\end{align}
where 
$V[d_{i},d^{*}_{i}]$ is the last term in Eq.~(\ref{eq:2020-05-31-01-33}),
which contains two or more sets of $d_{i},d_{i}^{*}$.
In the dual fermion system, 
the auxiliary particles (dual fermion) interact with each other regarding ${G}_{\rm 0dual}(k)$ as the non-interacting Green's function and $V$ as the interaction.
This resembles the Hubbard model, 
but 
differs in the points that 
the interaction exhibits frequency dependence and three or more particle terms.
If we can solve this dual fermion problem,
we can obtain the lattice quantities through the following relations 
which can be derived from definitions.
\begin{align}
  G^{-1}_{\rm lat}(k) %\nonumber \\ 
  =& 
  \Bigl(
    G_{\rm imp}(i\omega_{n}) + G_{\rm imp}(i\omega_{n})\Sigma_{\rm dual}(k) G_{\rm imp}(i\omega_{n})
  \Bigr)^{-1} \nonumber \\ 
  &\hspace{40pt}+ 
  \Delta(i\omega_{n}) - \epsilon_{\bm{k}},
  \label{eq:2020-05-31-11-50}
\end{align}
where 
$\Sigma_{\rm dual}$ is the self-energy in the dual fermion system.
\begin{align}
  F_{\rm lat}(k,k',q)
  =&
  L(k)L(k+q)F_{\rm dual}(k,k',q) R(k')R(k'+q),
  \label{eq:2020-05-31-12-00} \\
  L(k) =& G^{-1}_{\rm lat}(k)\Bigl( \Delta(i\omega_{n}) - \epsilon_{\bm{k}} \Bigr)^{-1} G_{\rm imp}^{-1}(i\omega_{n}) G_{\rm dual}(k),
  \label{eq:2020-05-31-12-01} \\
  R(k) =& G_{\rm dual}(k)G_{\rm imp}^{-1}(i\omega_{n}) \Bigl( \Delta(i\omega_{n}) - \epsilon_{\bm{k}} \Bigr)^{-1} G_{\rm lat}^{-1}(k),
  \label{eq:2020-05-31-12-02} 
\end{align}
where $G_{\rm dual}$ and $F_{\rm dual}$ are the Green's function and the full vertex in the dual fermion system, respectively.

\

\subsection{Ladder approximation}
Finally,
we will explain the ladder approximation used in this study to solve the dual fermion problem.
This corresponds to the fluctuation exchange (FLEX) approximation in the original lattice model 
and 
is an approximation which gives the leading correction to the DMFT from the perspective of $1/d$ expansion~\cite{PhysRevB.90.235132}.
The details of the approximation are as follows.

\begin{figure}[h]
  \centering
  \subfigure[self-energy]
  {\includegraphics[width=25mm,clip]{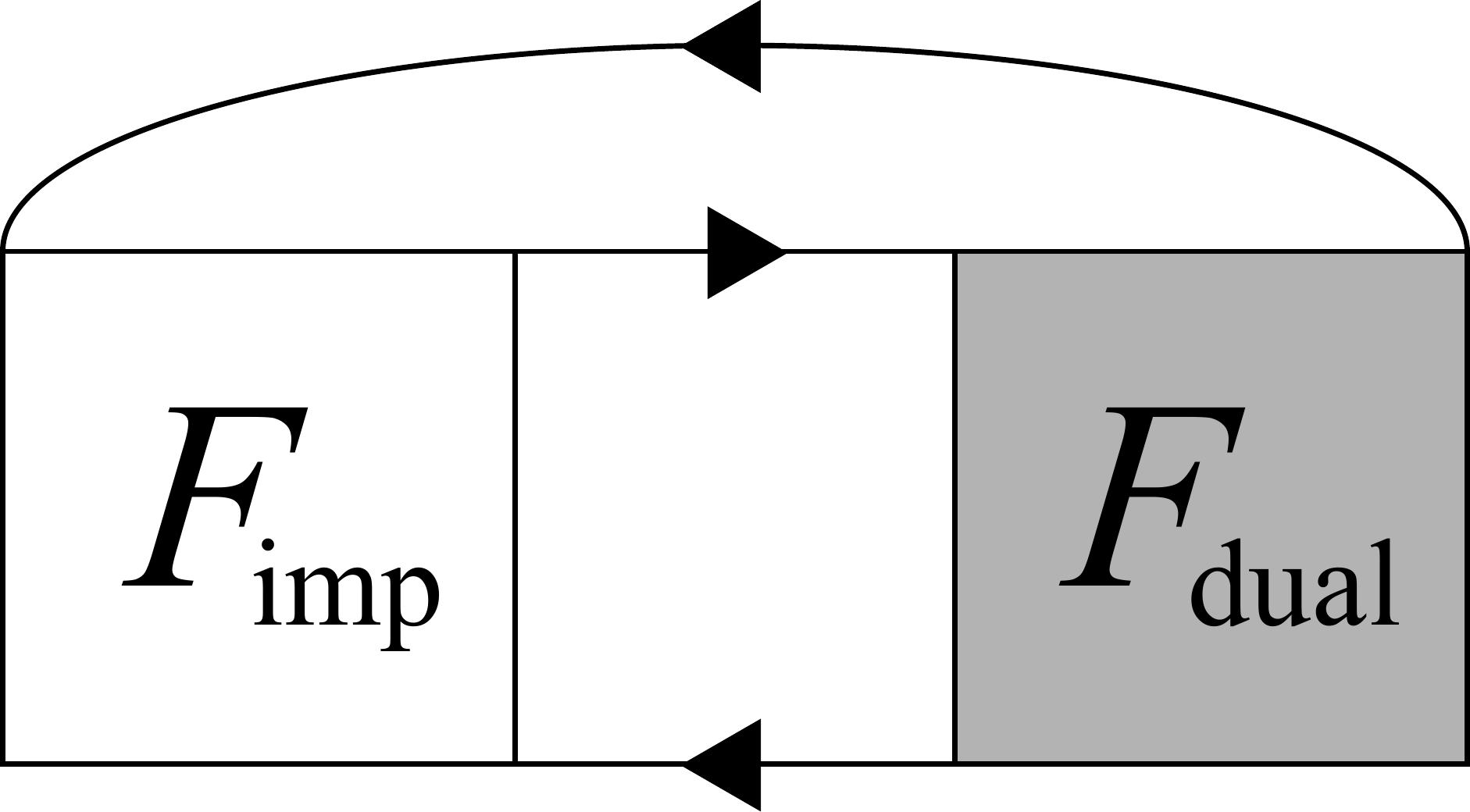}} \\
  \subfigure[Bethe-Salpeter equation]
  {\includegraphics[width=70mm,clip]{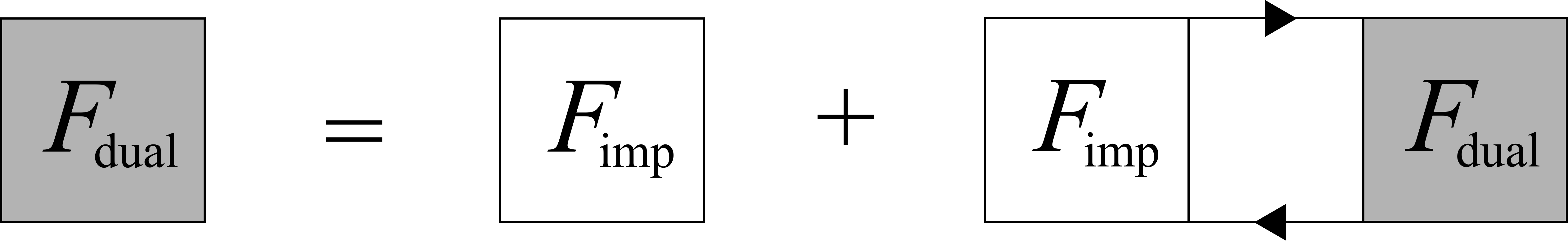}} 
  \caption{
    The diagrammatic representation of (a)~the self-energy and (b)~the Bethe-Salpeter equation in the dual fermion system in the ladder approximation.
    The bare interaction in the dual fermion system corresponds to the full vertex in the impurity system.
  }
  \label{fig:2020-06-11-19-27}
\end{figure}

First,
we ignore the three or more particle terms in $V[d_{i},d^{*}_{i}]$ of
Eq.~(\ref{eq:2020-05-31-02-48}), 
namely, 
\begin{align} 
  V[d_{i},d^{*}_{i}] 
  =&
  \dfrac{1}{4} \sum_{nn'm} F_{\rm imp}(i\omega_{n},i\omega_{n'},i\nu_{m})d^{*}_{\omega i \sigma_{1}}d^{*}_{\omega'+\nu i \sigma_{2}}d_{\omega' i \sigma_{3}}d_{\omega+\nu i \sigma_{4}}. 
  \label{eq:2020-06-11-15-48}
\end{align}
This enables us to apply the ordinary diagram expansion method
just by the replacement 
$U \to F_{\rm imp}$.
Further, 
assuming the 
SU(2) symmetric case,
we employ the approximation in which only the ladder diagrams depicted in 
Fig.~\ref{fig:2020-06-11-19-27} are taken into account. 
Then, 
the self-energy in the dual fermion system is obtained as follows:
\begin{align}
  {\Sigma}_{\rm dual}(k)
  =&
  \dfrac{1}{4}\sum_{q} \Bigl[
    V_{\rm c}(i\omega_{n},i\omega_{n'},i\nu_{m}, \bm{q}) 
    \nonumber \\
    &\hspace{20pt}+ 3V_{\rm s}(i\omega_{n},i\omega_{n'},i\nu_{m}, \bm{q})
  \Bigr]{G}_{\rm dual}(k+q),
  \label{eq:2020-06-11-16-06} 
\end{align}
\begin{align}
  V_{r}&(i\omega_{n},i\omega_{n'},i\nu_{m}, \bm{q})  \nonumber \\
  =&
  T \sum_{n''}
  F_{{\rm imp},r} (i\omega_{n},i\omega_{n''},i\nu_{m})
  {\chi}_{\rm 0dual}(i\omega_{n''},i\nu_{m}, \bm{q}) \nonumber \\ 
  &\hspace{0pt} \times 
  \Bigl[
    2F_{{\rm dual},r}(i\omega_{n''},i\omega_{n'},i\nu_{m}, \bm{q}) 
    -  
    F_{{\rm imp},r} (i\omega_{n''},i\omega_{n'},i\nu_{m})
  \Bigr],
  \label{eq:2020-06-11-16-10} 
\end{align}
\begin{align}
  F_{{\rm dual},r}& (i\omega_{n},i\omega_{n'},i\nu_{m}, \bm{q}) \nonumber \\
  =&
  F_{{\rm imp},r}(i\omega_{n},i\omega_{n'},i\nu_{m}) \nonumber \\
  -&
  \sum_{n''}
  F_{{\rm imp},r}(i\omega_{n},i\omega_{n''},i\nu_{m}) 
  \nonumber \\
  &\times {\chi}_{\rm 0dual}(i\omega_{n''},i\nu_{m}, \bm{q}) 
  F_{{\rm dual},r} (i\omega_{n''},i\omega_{n'},i\nu_{m}, \bm{q}),
  \label{eq:2020-06-11-16-34} 
\end{align}
where 
\begin{align}
  {\chi}_{\rm 0dual}(i\omega_{n},i\nu_{m},\bm{q}) 
  =&
  -\dfrac{1}{N_{\bm{k}}}\sum_{\bm{k}} {G}_{\rm dual}(k) {G}_{\rm dual}(k+q).
  \label{eq:2020-06-11-16-35}
\end{align}
Substituting this dual self-energy to 
Eq.~(\ref{eq:2020-05-31-11-50}), 
we can obtain the Green's function in the original lattice system.

\section{IPT + parquet method} \label{sec:2021-07-26-21-14}
Here, we briefly introduce the IPT + parquet method developed in Ref.~\cite{PhysRevB.104.035160}. 
In the IPT + parquet method, the correlation part of the self-energy $\hat{\Sigma}^{\rm CR}$ is obtained as 
\begin{align}
  \hat{\Sigma}^{\rm CR}_{\rm IPT + parquet}(\omega_{n}) =& [\hat{I} - \hat{B}\hat{\Sigma}^{\rm CR}_{0}(\omega_{n})]^{-1}\hat{A} \hat{\Sigma}^{\rm CR}_{0}(\omega_{n}), 
  \label{eq:2020-08-28-02-01} 
\end{align}
\begin{align} 
  {\Sigma}^{\rm CR}_{0,\alpha\beta}(\omega_{n})
  =&
  T^{2}\sum_{\gamma\lambda}\sum_{\omega_{n}'\nu_{m}}
  [\hat{F}_{0}(\omega_{n},\omega_{n'},\nu_{m}) 
    \hat{\chi}_{0}(\omega_{n'},\nu_{m})
  \hat{U} ]_{\alpha\gamma\beta\lambda} \nonumber \\
  &\hspace{30pt}\times G_{0,\gamma\lambda}(\omega_{n}+\nu_{m}),
  \label{eq:2020-06-04-14-05} \\
  &\chi_{0,\alpha\beta\gamma\lambda}(\omega_{n},\nu_{m}) = -G_{0,\alpha\gamma}(\omega_{n})G_{0,\lambda\beta}(\omega_{n}+\nu_{m}),
  \label{eq:2021-06-29-00-25}\\
  &\hat{G}_{0}(\omega_{n}) = [i\omega_{n}\hat{I}+\hat{\mu}_{0}- \hat{\Delta}(\omega_{n}) - \hat{\Sigma}^{\rm HF} ]^{-1},
  \label{eq:2020-07-02-19-01} 
\end{align}  
\begin{align}
  \hat{F}_{0}&(\omega_{n},\omega_{n'},\nu_{m}) 
  \nonumber \\
  =&
  \hat{U} + \hat{\Phi}_{\rm ph}(\nu_{m}) + \hat{\Phi}_{\rm \overline{ph}}(\omega_{n}-\omega_{n'}) + \hat{\Phi}_{\rm pp}(\omega_{n}+\omega_{n'}+\nu_{m}),
  \label{eq:2020-08-28-01-28}
\end{align}  
where $\hat{F}_{0}$ is an approximate full vertex and is obtained by the simplified parquet method developed in Ref.~\cite{doi:10.1143/JPSJ.79.094707} and extended for multiband systems in Ref.~\cite{PhysRevB.104.035160}.
$\hat{\Sigma}^{\rm HF}$ is the Hartree-Fock term of the self-energy.
$\hat{\mu}_{0}$ is a diagonal matrix and is determined such that $n_{0\alpha}=n_{\alpha}$ is satisfied,
where $n_{0\alpha}$ and $n_{\alpha}$ are the band filling obtained by $G_{0\alpha\alpha}$ and $G_{\alpha\alpha}$, respectively.
The parameters $\hat{A}$ and $\hat{B}$ are determined as follows.
\begin{align}
  A_{\alpha\beta} =& \delta_{\alpha\beta} \label{eq:2020-06-04-16-20} \\
  B_{\alpha\beta} =& \delta_{\alpha\beta} \dfrac{ N_{\rm orbital}^{-1} \sum_{\gamma}U_{\alpha\alpha\gamma\gamma}(1-2n_{\gamma}) + \mu_{0\alpha\alpha}-\mu}{\sum_{\gamma}U_{\alpha\alpha\gamma\gamma}n_{0\gamma}(1-n_{0\gamma})U_{\gamma\gamma\alpha\alpha}}.
  \label{eq:2020-06-04-16-19} 
\end{align}
Comparing the self-energy of IPT + parquet in Eqs.~(\ref{eq:2020-08-28-02-01}) to  (\ref{eq:2020-08-28-01-28}) with the exact self-energy
\begin{align}
  {\Sigma}^{\rm exact}_{0,\alpha\beta}(\omega_{n})
  =&
  T^{2}\sum_{\gamma\lambda}\sum_{\omega_{n}'\nu_{m}}
  [\hat{F}(\omega_{n},\omega_{n'},\nu_{m}) 
    \hat{\chi}_{0}(\omega_{n'},\nu_{m})
  \hat{U} ]_{\alpha\gamma\beta\lambda} \nonumber \\
  &\hspace{30pt}\times G_{\gamma\lambda}(\omega_{n}+\nu_{m}),
  \label{eq:2021-08-16-00-29} \\
  &\chi_{0,\alpha\beta\gamma\lambda}(\omega_{n},\nu_{m}) = -G_{\alpha\gamma}(\omega_{n})G_{\lambda\beta}(\omega_{n}+\nu_{m}),
  \label{eq:2021-08-16-00-30}
\end{align}
we obtain the full vertex of IPT + parquet in the form of Eqs.~(\ref{eq:2021-04-14-21-16}) and (\ref{eq:2021-04-14-21-17}) 
with 
${C}_{1}(\omega_{n})=[{I}-\hat{B}{\Sigma}^{\rm CR}_{0}(\omega_{n})]^{-1}{A}$
and ${C}_{2}(i\omega_{n})={C}_{3}(i\omega_{n})={C}_{4}(i\omega_{n})={G}_{0}(i\omega_{n}){G}(i\omega_{n})^{-1}$.
\section{Conditions for the crossing symmetry}\label{sec:2021-08-12-15-01}
We can express the scattering processes with the particle-hole and particle-particle pairs using the full vertex as follows.
\begin{align}
  \text{ph} =& \dfrac{1}{4} F_{\alpha\beta\gamma\lambda} c^{\dagger}_{\alpha}c^{\dagger}_{\lambda}c_{\gamma}c_{\beta},
  \label{eq:2021-08-12-14-43} \\
  \text{pp} =& \dfrac{1}{4} F^{\rm pp}_{\alpha\beta\gamma\lambda} c^{\dagger}_{\alpha}c^{\dagger}_{\beta}c_{\gamma}c_{\lambda}.
  \label{eq:2021-08-12-14-44} 
\end{align}
The following relations arise from the commutation relation of the annihilation and creation operators.
\begin{align}
  F(D)=&-F(C), \label{eq:2021-08-12-14-46}\\
  F^{\rm pp}(D)=&-F^{\rm pp}(T),\label{eq:2021-08-12-14-47}\\
  F^{\rm pp}(D)=&-F(P)=F(X),\label{eq:2021-08-12-14-48}
\end{align}
where the notations $D,T,C,P,X$ are given in Eqs.~(\ref{eq:2020-05-10-14-49}) to (\ref{eq:2020-05-10-14-53}).
The relations given in Eqs.~(\ref{eq:2021-08-12-14-46}) to (\ref{eq:2021-08-12-14-48}) are called the crossing symmetry.
If we obtain $F_{0}$ in Eq.~(\ref{eq:2021-04-14-21-17}) by the simplified parquet method, $F_{0}$  satisfies the crossing symmetry. 
To keep the symmetry while multiplying $C$'s, we need the condition $C_{1}=C_{2}=C_{3}=C_{4}$.
This condition is satisfied in S2F and not satisfied in IPT + parquet. 
Hence, the crossing symmetry is kept in (an impurity solver) + S2F + EDF and is violated in IPT + parquet + EDF. 
The omission of $C_{1}$ and $C_{3}$ in the EDF calculation does not affect the crossing symmetry since we omit them only when calculating the self-energy in Eq.(23). 
The two-particle full vertex always keeps the crossing symmetry in (an impurity solver) + S2F + EDF.

\section{Unphysical result in EDF}\label{eq:2021-08-11-21-21} 
\begin{figure}[] 
  \centering
  {\includegraphics[clip,width=90mm]{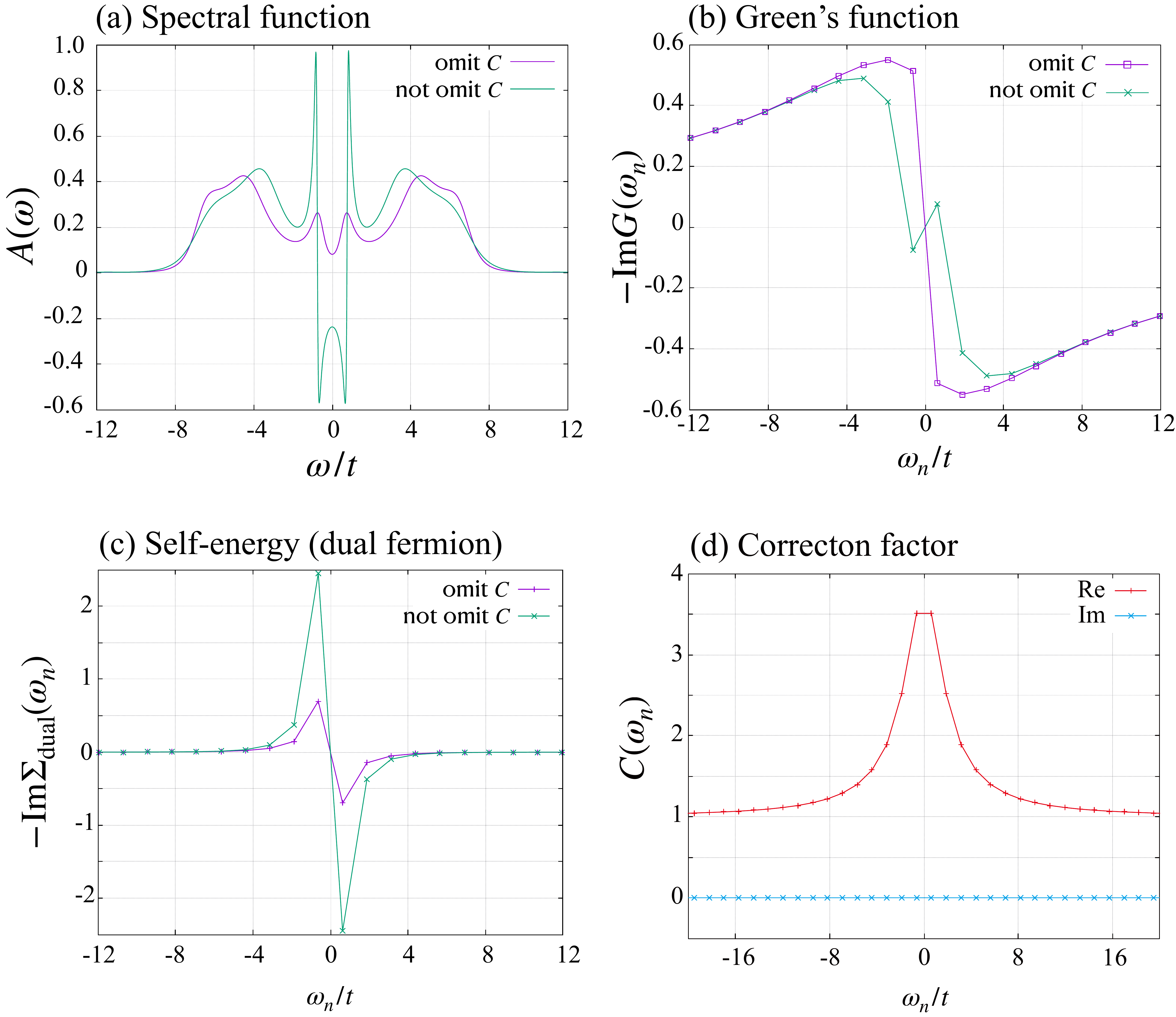}} 
  \caption{
    Quantities obtained by IPT + parquet + EDF on the single-orbital square lattice model.
    The interaction and temperature are $U/t=8$ and $T/t=0.2$, respectively.
    (a),(b),(c), and (d) show the Spectral function, the imaginary part of the Green's function in the lattice system, the imaginary part of the self-energy in the dual fermion system, and the correction factor, respectively.
    (a)-(c)~The violet~(green) line represents the case in which the correction factor $C_{3}$ is~(not) omitted in Eq.~(\ref{eq:2020-06-29-15-36}).
    (d) The red~(blue) line represents the real~(imaginary) part.
  } 
  \label{fig:2021-08-06-14-43}
\end{figure}

Here, 
we discuss the unphysical results caused by $C_{3}$ in the calculation of the self-energy of the dual fermion in Eq.~(\ref{eq:2020-06-29-15-36}) in IPT + parquet +EDF.
Figure~\ref{fig:2021-08-06-14-43} shows several quantities obtained by IPT + parquet + EDF (one-shot calculation) on the single-orbital square lattice model, which has only the nearest neighbor hopping and the same model as in Sec.~\ref{sec:2021-03-19-14-19}.
The interaction and temperature are $U/t=8$ and $T/t=0.2$, respectively.

Omitting (not omitting) $C_{3}$ in Eq.~(\ref{eq:2020-06-29-15-36}) gives the results depicted by the violet (green) line in Fig~\ref{fig:2021-08-06-14-43}~(a) to (c).
When we do not omit $C_{3}$, 
the spectral function has a negative value in the low frequency region and the causality is violated as in Fig.~\ref{fig:2021-08-06-14-43}~(a).
This is the unphysical result mentioned in Sec.~\ref{sec:2020-07-10-21-41}
and 
can be circumvented by omitting $C_{3}$.
This can be seen also in Fig.~\ref{fig:2021-08-06-14-43}~(b).
In Fig.~\ref{fig:2021-08-06-14-43}~(c),
we can see that the low-frequency part of the dual self-energy is increased by $C_{3}$ depicted in Fig.~\ref{fig:2021-08-06-14-43}~(d).
This overly suppress the low-energy part of the lattice Green's function in Eq.~(\ref{eq:2020-05-31-11-50})
and then
violate the causality as in Fig.~\ref{fig:2021-08-06-14-43}~(b).
Although we cannot be certain at present,
we speculate the possibility that the correction factor $C$ is overestimated as mentioned in Sec.~\ref{sec:2021-03-19-14-19}. 
Since the IPT + parquet results agrees very well with the CT-QMC results while underestimating the diagonal structure, 
the cross and central structures given by the correction factor $C$ may compensate the differences.
The same is true in S2F since we determine the $C$ which gives the cross and central structures to reproduce the self-energy obtained by CT-QMC while using the (underestimated) diagonal structure obtained by the simplified parquet method.
If the overestimation~(underestimation) of the cross and central~(diagonal) structures can be suppressed through the modification we suggest in Sec.~\ref{sec:2021-08-03-16-22},
the unphysical results in Fig.~\ref{fig:2021-08-06-14-43} will not appear and we may not have to omit $C$ in EDF.

\bibliographystyle{../../bibfiles/prb}
\bibliographystyle{apsrev4-1}
\bibliography{../../bibfiles/reference}

%\bibliography{Mizuno_paper2}

\end{document}